\documentclass[10pt,prd,reprint,aps,nofootinbib,nobibnotes,superscriptaddress,preprintnumbers]{revtex4-2}
\usepackage{graphicx,bm,amsmath,amssymb, mathtools,multirow, enumitem}
\graphicspath{{./suppression_theory_figs/}}
\usepackage[dvipsnames]{xcolor}

\usepackage[hidelinks]{hyperref}
\hypersetup{
  colorlinks   = true, %Colours links instead of ugly boxes
  urlcolor     = [RGB]{46,48,146}, %Colour for external hyperlinks
  linkcolor    = [RGB]{46,48,146}, %Colour of internal links
  citecolor    = [RGB]{46,48,146} %Colour of citations
}

\usepackage{braket,orcidlink}
\usepackage{relsize}

\newcommand{\EQ}{\begin{equation}}
\newcommand{\EN}{\end{equation}}
\newcommand{\EQA}{\begin{eqnarray}}
\newcommand{\ENA}{\end{eqnarray}}

{}
{}
{}

\usepackage[percent]{overpic}
\usepackage{placeins}
\allowdisplaybreaks

\begin{document}

\title{Finite-distance effects and scalar suppression in massive gravity}

\date{\today}
\author{Chris~Choi\,\orcidlink{0009-0005-2328-3044}}
\email{Contact author: minyeonc@andrew.cmu.edu}
\affiliation{McWilliams Center for Cosmology and Astrophysics and Department of Physics, \href{https://ror.org/05x2bcf33}{Carnegie Mellon University}, Pittsburgh, Pennsylvania 15213, USA}

\author{Tina~Kahniashvili\,\orcidlink{0000-0003-0217-9852}}
\email{Contact author: tinatin@andrew.cmu.edu}
\affiliation{McWilliams Center for Cosmology and Astrophysics and Department of Physics, \href{https://ror.org/05x2bcf33}{Carnegie Mellon University}, Pittsburgh, Pennsylvania 15213, USA}
\affiliation{School of Natural Sciences and Medicine, \href{https://ror.org/051qn8h41}{Ilia State University}, GE-0194 Tbilisi, Georgia}
\affiliation{Department of Theoretical Astrophysics and Cosmology, \href{https://ror.org/02gkgrd84}{Georgian National Astrophysical Observatory}, Tbilisi, GE-0179, Georgia}

\author{Barnab\'as~P\'oczos\,\orcidlink{0000-0002-2898-4705}}
\email{Contact author: bapoczos@andrew.cmu.edu}
\affiliation{Machine Learning Department, \href{https://ror.org/05x2bcf33}{Carnegie Mellon University}, Pittsburgh, Pennsylvania 15213, USA}

\begin{abstract}
Pulsar timing arrays (PTAs) probe the stochastic gravitational wave (GW) background through the angular cross-correlations of timing residuals. For a GW background in general relativity (GR), the expected overlap reduction function (ORF) is given by the Hellings--Downs (HD) correlation. We investigate how a nonzero graviton mass modifies this prediction through a massive dispersion relation and additional vector and scalar polarizations, producing a modified effective ORF. In the massive limit, the finite-distance pulsar terms survive angular averaging, multiplying the correlations of all three polarization sectors by a pulsar-dependent factor that can enhance, suppress, or reverse them. We adopt a tensor Earth–Earth normalization that preserves the continuous finite-distance effect. In the massless limit, we find that the vector ORF vanishes and the scalar ORF approaches a finite breathing mode. We propose the phenomenological prescription $\Omega_S/\Omega_T=A_S(m_g/\omega)^{n_S}$, with scalar amplitude parameter $A_S$ and suppression index $n_S>0$. These results provide a consistent framework for future PTA analyses of massive gravity.
\end{abstract}

\maketitle

\section{Introduction}\label{sec:intro}
Pulsar timing arrays (PTAs) search for gravitational waves (GWs) by monitoring the arrival times of electromagnetic pulses from millisecond pulsars over a period of many years \cite{Sazhin:1978myk, Detweiler:1979wn}. A passing GW perturbs the observed frequency of the pulse, inducing a redshift. The corresponding timing residuals are obtained from integrating this redshift, and can be correlated between different pulsars to search for a stochastic GW background (SGWB). For an isotropic background, the timing correlations between a pair of pulsars are described by polarization dependent GW energy density spectra and overlap reduction functions (ORFs), which generally depend on frequency, angular separation, and pulsar distances. In general relativity (GR), the tensor ORF for an SGWB reduces to the Hellings--Downs (HD) correlation \cite{Hellings:1983fr} in the infinite-distance approximation, making it dependent only on the separation angle between pulsars. Theories beyond GR can modify this correlation pattern in various ways, including its shape and dependence on other parameters, such as frequency or group velocity of the GW, $v_g$. PTA observations have also been used to constrain deviations from GR, including the presence of non-Einsteinian polarization modes \cite{NANOGrav:2021ini,NANOGrav:2023ygs}. PTA data sets are thus a rich avenue for testing alternative theories of gravity.

One theory of particular interest is massive gravity, in which the graviton has a nonzero mass, $m_g \neq 0$. In such a theory, a massive spin-$2$ field generically carries five polarization degrees of freedom: in addition to the two helicity-$\pm 2$ tensor modes of general relativity, it also includes two helicity-$\pm 1$ vector modes and one helicity-$0$ scalar mode. The effects of massive gravity are most naturally relevant on scales comparable to the graviton Compton wavelength. A graviton mass of cosmological order, $m_g \sim H_0$, where $H_0$ is the present-day Hubble parameter, corresponds to a Compton wavelength comparable to the Hubble scale and can therefore affect late-time cosmology \cite{DAmico:2011eto}. Certain branches of massive gravity admit self-accelerating solutions, potentially providing an alternative to a cosmological constant \cite{Koyama:2011wx,Gratia:2012wt}. However, constructing a consistent theory of massive gravity is highly nontrivial: nonlinear extensions of Fierz--Pauli massive gravity suffer from a pathological sixth degree of freedom, the Boulware--Deser ghost \cite{Boulware:1972yco}. More recently, the construction of ghost-free nonlinear massive gravity by de Rham, Gabadadze, and Tolley (dRGT) \cite{deRham:2010ik,deRham:2010kj}, together with its extension to bigravity \cite{Hassan:2011zd}, have established a consistent framework for studying massive spin-2 fields (see Refs.~\cite{Hinterbichler:2011tt,deRham:2014zqa,deRham:2016nuf,deRham:2023ngf} for reviews). This framework has since motivated a broad literature on cosmological backgrounds and self-accelerating solutions in massive gravity and bigravity \cite{DAmico:2011eto, Gratia:2012wt, Gumrukcuoglu:2012aa, Maeda:2013bha, Akrami:2013pna, Koyama:2011wx, Tasinato:2012ze, Koennig:2014ods}. Other work has studied perturbative stability, extensions to cosmological settings, and the viability of different branches of solutions \cite{DeFelice:2013awa,Gumrukcuoglu:2013nza,DeFelice:2013bxa,Cusin:2016ytz, Gumrukcuoglu:2016hic, Kenna-Allison:2019tbu, Comelli:2013tja, Heisenberg:2024uwq, Wood:2025ujj}, while observational analyses have placed constraints on bimetric and massive-gravity cosmologies \cite{Lambiase:2012fv,Hogas:2019ywm,Hogas:2021fmr,Hogas:2021lns,Hogas:2021saw,Smirnov:2025yru, Hogas:2025ahb}.

Pulsar timing arrays provide a way to test a nonzero graviton mass through its effects on GW propagation and polarization content. Previous PTA studies have explored subluminal GW propagation \cite{Lee:2010cg, Wu:2023rib, Bi:2023ewq, Wang:2023div, Wu:2023pbt}, modifications to the HD correlation \cite{Liang:2021bct, Arjona:2024cex, Cordes:2024oem, Domenech:2024pow}, and non-Einsteinian polarizations \cite{NANOGrav:2021ini, NANOGrav:2023ygs, Hu:2025rlz, Bernardo:2023zna, Chamberlin:2011ev, Hu:2026iln}. These effects enter the observed timing correlations through both the pulsar response to each polarization and the relative power carried by the different sectors.

At the level of the pulsar response, an important consideration is whether the pulsar-dependent terms can be neglected. These terms are often averaged over or neglected in ORF calculations in PTA analyses \cite{Hellings:1983fr, Anholm:2008wy, Romano:2023zhb, Allen:2022ksj, Liang:2021bct, Wu:2023rib}. Although this approximation is generally appropriate for the GR tensor modes, it does not necessarily hold across the propagation regimes and polarization sectors relevant to massive gravity. Recent studies have examined finite-distance effects and the limitations of the infinite-distance approximation \cite{Cordes:2024oem, Domenech:2024pow, Bernardo:2022rif, Liang:2024mex, Hu:2026iln}. These considerations motivate retaining the full finite-distance response and adopting a normalization that remains consistent when the pulsar terms are included.

The relative sector energy densities are particularly important near the massless limit. As we demonstrate below, the scalar ORF approaches a finite and nonzero correlation as $m_g \to 0$, whereas the vector ORF vanishes. A model intended to recover GR must also specify how the scalar energy density behaves in this limit. We introduce a phenomenological scalar energy-density prescription. We establish sufficient conditions for uniform scalar decoupling and recovery of the finite-distance GR tensor correlation. This work develops the theoretical framework and studies its phenomenology; parameter estimation and observational constraints using PTA data are left for future work.

This paper is organized as follows. In Sec.~\ref{sec:formalism}, we introduce the formalism for the analysis of the SGWB in massive gravity. In Sec.~\ref{sec:finite_dist}, we study the finite-distance effect and the massive and massless limits. In Sec.~\ref{sec:screen}, we introduce a prescription for suppressing the scalar contribution. In Sec.~\ref{sec:conclusion}, we discuss our findings and the possibility of future exploration. In what follows, we use natural units, setting $c=\hbar=k_B=1$, and we use the $+---$ metric convention.

\section{Formalism}\label{sec:formalism}
\subsection{Dispersion and polarization conventions}
\label{subsec:ghost_free}
When the region of observation is small compared with the background curvature scale, we may express the metric perturbation $h_{\mu\nu}$ as a plane wave \cite{Isi:2018miq, Liang:2021bct}
\begin{equation}\label{eq:planewave}
    h_{\mu\nu}(x) = \frac{1}{2\pi}\int d^4 k  \frac{2\delta(\omega^2 - |\boldsymbol{k}|^2 - m_g^2)}{|\boldsymbol{k}|} e^{ikx} h_{\mu\nu}(k) ,
\end{equation}
where $\omega=2\pi f$ is the angular frequency and $\boldsymbol{k}$ is the spatial wave vector. We adopt the Fourier normalization and integration measure of Ref.~\cite{Liang:2021bct}. In this convention, the factor
$2/|\boldsymbol{k}|$ is chosen so that after performing the integral, the measure reduces to $df\,d^2\hat{\boldsymbol{\Omega}}$. We have also used the notation $kx \equiv k_\mu x^\mu$, where $x^\mu=(t,\boldsymbol{x})$ is the spacetime coordinate at which the perturbation is evaluated and $k^\mu = (\omega, \boldsymbol{k})$ is the wave four-vector of the metric perturbation. The measure $d^4k$ denotes integration over all components of $k^\mu$, while the delta function restricts the modes to the mass shell via the dispersion relation,
\begin{equation}\label{eq:dispersion_relation}
    \omega^2 = |\boldsymbol{k}|^2 + m_g^2 .
\end{equation}
In general massive-gravity cosmologies, the dispersion relation does not necessarily reduce to this form. In a Friedmann--Lemaître--Robertson--Walker background, the equation of motion can contain a time-dependent propagation speed and an effective mass associated with GWs that need not equal the graviton mass parameter $m_g$, such as in extensions of quasi-dilaton theories in massive gravity \cite{DeFelice:2013tsa}. As an example, after neglecting terms related to cosmological background evolution, the dispersion relation for the propagating tensor modes takes the form \cite{Gumrukcuoglu:2012wt}
\begin{equation}\label{eq:general_tensor_dispersion}
    \omega^2 = c_g^2(t)\frac{|\boldsymbol{k}|^2 + 2K}{a(t)^2} + M_{\rm GW}^2(t) ,
\end{equation}
where $M_{\rm GW}(t)$ is the effective mass of the tensor modes, $a(t)$ is the scale factor, $K$ is the spatial curvature, and $c_g(t)$ is the tensor sound speed, whose square multiplies the spatial gradient term. The distinction between $M_{\rm GW}$ and $m_g$ is explicit in quasidilaton and extended quasidilaton massive gravity, where the effective GW mass can differ from $m_g$ \cite{Kahniashvili:2014wua}. In the present analysis, we assume $a=1$, $c_g=1$, a flat background ($K = 0$), and $M_{\rm GW} = m_g$, leading to the dispersion relation in Eq.~\ref{eq:dispersion_relation}.

Unlike in GR, the frequency cannot be arbitrarily small: the dispersion relation requires $\omega\ge m_g$, with the minimum value $\omega_{\rm min}=m_g$ reached when $|\boldsymbol{k}|=0$. We define the group velocity of the GW
\begin{equation}\label{eq:group_velocity}
    v_g \equiv \frac{|\boldsymbol{k}|}{\omega}  = \sqrt{1 - \frac{m_g^2}{\omega^2}}.
\end{equation}
The physical group velocity is obtained from the full dispersion relation and is not in general equal to $c_g$; the two coincide only when the $c_g^2|\boldsymbol{k}|^2$ term dominates the massive term.

We express $h_{\mu \nu}$ as
\begin{equation}\label{eq:h_munu}
    h_{\mu\nu}(k) = \sum_{I\in\{T,V,S\}}\sum_{i\in I} h^{(i)} (k) \epsilon_{\mu\nu}^{(i)} ,
\end{equation}
where we have grouped the helicity states into their sectors, $T = \{+2,-2\}$, $V = \{+1,-1\}$, and $S = \{0\}$, following standard polarization decompositions \cite{Isi:2018miq, Chamberlin:2011ev, Liang:2021bct}. Polarization tensors $\epsilon_{\mu\nu}^{(i)}$ can be constructed using a basis of polarization vectors adapted from $\hat{\boldsymbol{\Omega}}$, the propagation direction of the GW, which in general is expressed as $\hat{\boldsymbol{\Omega}}\equiv (\sin\theta\cos\phi,\sin\theta\sin\phi,\cos\theta)$. 

We use the massive spin-1 polarization vectors \cite{Liang:2021bct}
\begin{equation}\label{eq:polarization_vectors}
\begin{aligned}
    \epsilon^\mu_{\pm}(k) &= \frac{1}{\sqrt{2}}
\begin{pmatrix}
0 \\
\cos\theta\cos\phi \mp i\sin\phi \\
\cos\theta\sin\phi \pm i\cos\phi \\
-\sin\theta
\end{pmatrix}, \\
\epsilon^\mu_{0}(k)
&=
\begin{pmatrix}
\frac{|\boldsymbol{k}|}{m_g} \\
\frac{\omega}{m_g} \sin\theta\cos\phi \\
\frac{\omega}{m_g}\sin\theta\sin\phi \\
\frac{\omega}{m_g}\cos\theta
\end{pmatrix}.
\end{aligned}
\end{equation}
These vectors are transverse to $k_\mu$, satisfying $k_\mu\epsilon^\mu_{n}=0$. We construct the massive spin-2 polarization tensors from the spin-1 polarization vectors via Clebsch--Gordan coefficients \cite{Cohen-Tannoudji:1986qmv, Varshalovich:1988ifq},
\begin{equation}
\begin{aligned}
\epsilon_{\mu\nu}^{(+2)} &= \epsilon_\mu^{+}\epsilon_\nu^{+}, \\
\epsilon_{\mu\nu}^{(+1)} &= \frac{1}{\sqrt{2}}
\left(\epsilon_\mu^{+}\epsilon_\nu^{0}+\epsilon_\mu^{0}\epsilon_\nu^{+}\right), \\
\epsilon_{\mu\nu}^{(0)} &= \frac{1}{\sqrt{6}}
\left(\epsilon_\mu^{+}\epsilon_\nu^{-}+\epsilon_\mu^{-}\epsilon_\nu^{+}
      -2\,\epsilon_\mu^{0}\epsilon_\nu^{0}\right), \\
\epsilon_{\mu\nu}^{(-1)} &= \frac{1}{\sqrt{2}}
\left(\epsilon_\mu^{-}\epsilon_\nu^{0}+\epsilon_\mu^{0}\epsilon_\nu^{-}\right), \\
\epsilon_{\mu\nu}^{(-2)} &= \epsilon_\mu^{-}\epsilon_\nu^{-} \ .
\end{aligned}
\end{equation}
These tensors are transverse $k^\mu\epsilon_{\mu\nu}^{(i)}=0$ and traceless $\eta^{\mu\nu}\epsilon_{\mu\nu}^{(i)}=0$ for each helicity $i$. Transversality follows from that of the polarization vectors. In our phase convention, $\eta^{\mu\nu}\epsilon_\mu^+\epsilon_\nu^- = \eta^{\mu\nu} \epsilon_\mu^0 \epsilon_\nu^0 = -1$, so the relative minus sign in the helicity-$0$ tensor ensures tracelessness. Additionally, they are unit normalized, $\epsilon_{\mu\nu}^{(i)*}\epsilon^{(j)\mu\nu} = \delta_{ij}$.

\subsection{Pulsar timing array response}
We now consider the response of a pulsar signal to a massive GW propagating along the direction $\hat{\boldsymbol{\Omega}}$. The GW perturbs the spacetime along the line of sight of the pulsar signals, originally emitted with a constant frequency $\nu_0$, as they travel from their source pulsars to our telescopes. The wave-vector of the photon pulse is explicitly $\hat{p}_r^\mu = (1, - \hat{\boldsymbol{p}}_r)$, where $\hat{\boldsymbol{p}}_r$ points from Earth to the pulsar $r$. In the barycenter reference frame of the solar system, the frequency shift is characterized by the redshift $z(t)\equiv(\nu_0-\nu(t))/\nu_0$, following the PTA convention of Refs.~\cite{Anholm:2008wy,Liang:2021bct}. PTA collaborations directly measure the times of arrival of pulsar signals, which correspond to the integral of $z(t)$ \cite{Anholm:2008wy, Dahal:2020}, otherwise known as the timing residual. In Fourier space, the redshift for signals emitted by pulsar $r$, located a distance $L_r$ away, can be written as
\begin{equation}\label{eq:redshift}
\begin{aligned}
\tilde{z}_r(f,\hat{\boldsymbol{\Omega}})
&= \sum_{I\in\{T,V,S\}}\sum_{i\in I}
\left[
e^{-i2\pi f L_r\left(1+v_g\hat{\boldsymbol{\Omega}}\cdot\hat{\boldsymbol{p}}_r\right)}
-1
\right] \\
&\quad \times
h^{(i)}\left(f,v_g\hat{\boldsymbol{\Omega}}\right)
F_r^{(i)}(\hat{\boldsymbol{\Omega}}) ,
\end{aligned}
\end{equation}
where the tilde on the redshift indicates that it is a quantity in Fourier space, and the bracketed factor is the difference between the pulsar term and the Earth term, which comes from the retarded waveforms. $F^{(i)}_r(\hat{\boldsymbol{\Omega}})$ is the geometrical projection factor, which in our decomposition is equal to \cite{Liang:2021bct}
\begin{equation}\label{eq:receiving}
        F^{(i)}_r(\hat{\boldsymbol{\Omega}})\! =\! -\frac{\hat{p}_r^\mu \hat{p}_r^\nu}{2(1+v_g \hat{\boldsymbol{\Omega}} \cdot \hat{\boldsymbol{p}}_r)} \epsilon_{\mu \nu}^{(i)}+\hat{p}_r^\mu \epsilon_{0 \mu}^{(i)}-\frac{1}{2} \epsilon_{00}^{(i)}(1+v_g \hat{\boldsymbol{\Omega}} \cdot \hat{\boldsymbol{p}}_r).
\end{equation}
The second and third terms come from the time component of the metric perturbation in the redshift response. These terms vanish in the synchronous gauge, as seen in Ref.~\cite{Chamberlin:2011ev}, but they are nonzero in the unitary gauge that we adopt.

To detect an SGWB, we look at the cross-correlation of the residuals across all pulsars, which we can obtain via the two-point correlation of the total redshift. We assume that there are two pulsars $a$ and $b$, located at distances $L_a$ and $L_b$ from the solar-system barycenter in directions $\hat{\boldsymbol{p}}_a$ and $\hat{\boldsymbol{p}}_b$. We assume that pulsar $a$ is situated along the $z$-axis and pulsar $b$ is located in the $x$--$z$ plane at an angle $\xi_{ab}$ from pulsar $a$. Then, the pulsar directions are explicitly
\begin{equation}
\hat{\boldsymbol p}_a=(0,0,1), \qquad \hat{\boldsymbol p}_b=(\sin\xi_{ab},0,\cos\xi_{ab}).
\end{equation}

Assuming that the SGWB is stationary and isotropic, the corresponding two-point correlation function of the sky-integrated redshift, $\tilde{z}_a(f)\equiv\int d^2\hat{\boldsymbol{\Omega}}\,z_a(f,\hat{\boldsymbol{\Omega}})$, is \cite{Liang:2021bct,Anholm:2008wy}
\begin{equation}\label{eq:two_point_z}
\begin{aligned}
\left\langle
\tilde z_a^*(f)\tilde z_b(f')
\right\rangle
&=
\frac{3H_0^2}{8\pi^3}
\delta(f-f')|f|^{-3} \\
&\quad\times
\sum_{I \in\{T,V,S\}} \sum_{i \in I} \Omega_i(|f|)
\frac{\widetilde{\Gamma}_{ab}^{i}(\xi_{ab},|f|)}{\beta_i(f)} .
\end{aligned}
\end{equation}
Here $\Omega_i$ is the dimensionless energy density per logarithmic frequency for each helicity $i$, $\widetilde{\Gamma}_{ab}^{i}$ is the corresponding normalized ORF, and $\beta_i$ is the normalization factor. The raw signal $\left\langle \tilde z_a^*(f)\tilde z_b(f') \right\rangle$ itself is independent of normalization convention, since the normalized ORF is being divided by its normalization. The factor $\delta(f-f')$ follows from stationarity, and isotropy removes dependence on the absolute orientation of the pulsar pair via rotational symmetry, leaving the angular dependence only through $\xi_{ab}$.

Within the tensor and vector sectors, we assume equal power in the two helicity modes. This follows from taking the stochastic background to be parity invariant, so that there is no net circular polarization in either the tensor or vector sector. We also assume that there is no cross-correlation between distinct helicities or sectors, motivated by the fact that the tensor, vector, and scalar modes decouple with an appropriate gauge-fixing term, allowing the quadratic kinetic terms to be diagonalized \cite{Hinterbichler:2011tt}. We can write the dimensionless energy density and ORF for the sectors themselves, with $\Omega_I = \sum_{i \in I} \Omega_i$ and unnormalized ORF $\Gamma_{ab}^I = \sum_{i \in I} \Gamma_{ab}^i$. Thus, with the equal power assumption, $\Omega_{+2}=\Omega_{-2}=\Omega_T/2$ and $\Omega_{+1}=\Omega_{-1}=\Omega_V/2$, while $\Omega_0=\Omega_S$. The signal is proportional to
\begin{equation}
\begin{aligned}
\left\langle\widetilde z_a^*(f)\widetilde z_b(f')\right\rangle
&\propto
\frac{\Omega_T}{2}\Gamma_{ab}^{+2}
+\frac{\Omega_T}{2}\Gamma_{ab}^{-2}
+\frac{\Omega_V}{2}\Gamma_{ab}^{+1}
\\ &
+\frac{\Omega_V}{2}\Gamma_{ab}^{-1}
+\Omega_S\Gamma_{ab}^{0}
\\
&=\frac{\Omega_T}{2}
\Gamma^T_{ab}
+\frac{\Omega_V}{2}\Gamma^V_{ab}+\Omega_S\Gamma^S_{ab},
\end{aligned}
\end{equation}
leaving the scalar contribution with a relative factor of 2 compared to the tensor and vector contributions.

Following Ref.~\cite{Liang:2021bct}, we define the ORF for each polarization sector $I$ as
\begin{equation}\label{eq:orf_type}
\begin{aligned}
\widetilde{\Gamma}_{ab}^{I}(\xi_{ab},f)
&\equiv \beta_I \Gamma_{ab}^I \\ &= \beta_I \operatorname{Re}\!\Bigg[ \sum_{i\in I}
\int_{S^2} d^2\hat{\boldsymbol{\Omega}}\,
\left[
e^{i2\pi f L_a(1+v_g\hat{\boldsymbol{\Omega}}\cdot\hat{\boldsymbol{p}}_a)} -1
\right] \\
& \times
\left[ e^{-i2\pi f L_b(1+v_g\hat{\boldsymbol{\Omega}}\cdot\hat{\boldsymbol{p}}_b)}
-1 \right] F_a^{(i)\,*}(\hat{\boldsymbol{\Omega}}) F_b^{(i)}(\hat{\boldsymbol{\Omega}}) \Bigg] .
\end{aligned}
\end{equation}
We note that the phases have opposite signs because the redshift from pulsar $a$ is complex conjugated in $\langle\tilde z_a^*(f)\tilde z_b(f')\rangle$, while the redshift from pulsar $b$ is not. If we cannot distinguish among the different polarization modes, we observe the angular correlation as an effective ORF normalized relative to the tensor sector \cite{Liang:2021bct}, which has the following form 
\begin{equation}\label{eq:eff_orf_2}
    \widetilde{\Gamma}_{ab}^{\rm eff}(\xi_{ab},f) = \widetilde{\Gamma}_{ab}^{T}+  \frac{\Omega_V}{\Omega_T} \frac{\beta_{T}}{\beta_{V}} \widetilde{\Gamma}_{ab}^{V}+ 2 \frac{\Omega_S}{\Omega_T} \frac{\beta_{T}}{\beta_{S}}\widetilde{\Gamma}_{ab}^{S}.
\end{equation}
Here, we notice that the ratios $\beta_T/\beta_{V}$ and $\beta_T/\beta_{S}$ cause the normalization factors for the vector and scalar modes to cancel out in their ORFs. We may factor $\beta_T$ out of the expression and equivalently write the effective ORF as
\begin{equation}\label{eq:eff_orf_common_beta}
\widetilde{\Gamma}_{ab}^{\rm eff} =
\beta_T \left[\Gamma_{ab}^{T} + \frac{\Omega_V}{\Omega_T}\Gamma_{ab}^{V} + 2 \frac{\Omega_S}{\Omega_T} \Gamma_{ab}^{S}
\right].
\end{equation}
Alternatively, if the combined correlation were interpreted as an effective vector or scalar signal, we could analogously factor out $\beta_V$ or $\beta_S$, with the relative energy densities expressed using $\Omega_V$ or $\Omega_S$ in the denominator, respectively.

\section{Finite-distance effect}\label{sec:finite_dist}
The finite-distance effect enters through the difference between the metric perturbation evaluated at the pulsar and at the Earth. Following Ref.~\cite{Liang:2021bct}, we define the difference in metric perturbation as
\begin{equation}
\Delta h_{\alpha\beta}
\equiv
h_{\alpha\beta}
\left(t_p,v_g\hat{\boldsymbol{\Omega}}\right)
-
h_{\alpha\beta}
\left(t_e,v_g\hat{\boldsymbol{\Omega}}\right).
\end{equation}
Using $\mathbf{x}_e=0,\ \mathbf{x}_p=L_r\hat{\boldsymbol{p}}_r,\ t_p=t_e-L_r$, and writing the metric perturbation as a plane wave, this difference becomes
\begin{equation}
\begin{aligned}
\Delta h_{\mu\nu}
&=
\int_{-\infty}^{\infty} df\,
e^{i2\pi f t_e}
\left[
e^{-i2\pi fL_r
\left(1+v_g\hat{\boldsymbol{\Omega}}\cdot\hat{\boldsymbol{p}}_r\right)}
-1
\right]  \\
&\quad \times
\sum_i
h^{(i)}
\left(f,v_g\hat{\boldsymbol{\Omega}}\right)
\epsilon_{\mu\nu}^{(i)}(\hat{\boldsymbol{\Omega}}).
\end{aligned}
\end{equation}
The factor in square brackets contains both the pulsar term and the Earth term, which we notate as
\begin{equation}\label{eq:P_r}
    \mathcal{P}_{r}(\hat{\boldsymbol{\Omega}}) \equiv e^{-i2\pi fL_r \left(1+v_g\hat{\boldsymbol{\Omega}}\cdot\hat{\boldsymbol{p}}_r\right)}, \qquad \mathcal{P}_{0}=1.
\end{equation}
The ORFs therefore contain the product
\begin{equation}\label{eq:finite_distance_full}
\begin{aligned}
    \left(\mathcal{P}_{a}^{*}-\mathcal{P}_{0}^*\right)
    \left(\mathcal{P}_{b}-\mathcal{P}_{0}\right)
    &=\left(\mathcal{P}_{a}^{*}-1\right)
    \left(\mathcal{P}_{b}-1\right) \\ &=
    \mathcal{P}_{a}^{*}\mathcal{P}_{b}
    -\mathcal{P}_{a}^{*}
    -\mathcal{P}_{b}
    +1 .
\end{aligned}
\end{equation}
The four terms in Eq.~\eqref{eq:finite_distance_full} are the pulsar--pulsar, pulsar--Earth, Earth--pulsar, and Earth--Earth contributions, respectively. We refer to the first three collectively as the finite-distance terms because they retain the dependence on the pulsar distances through $\mathcal{P}^*_a$ and $\mathcal{P}_b$. We use the term ``finite-distance effect'' for the combined modification to the ORF that results from the retention of these three terms. In the usual large-$fL$ limit, their rapidly oscillating phases in the angular integral cause them to average away, but this behavior is not uniform as $v_g \to 0$ or as the two pulsar phases become coherent.

It is useful to assign these four terms their own ORFs. For each sector $I\in\{T,V,S\}$, we write
\begin{equation}
\Gamma_{ab}^{I} = \Gamma_{ab}^{a^*b,I} -\Gamma_{ab}^{a^*0,I} -\Gamma_{ab}^{0b,I} +\Gamma_{ab}^{00,I},
\label{eq:sector_EP_decomp}
\end{equation}
where the superscript $a^*$ denotes that the term involves pulsar $a$ and is complex conjugated, the superscript $b$ denotes the term involves pulsar $b$, and the superscript 0 denotes the Earth term. Explicitly, each term is defined as
\begin{equation}
\Gamma_{ab}^{x^*y,I} =\operatorname{Re}\left[ \sum_{i\in I}\int_{S^2}d^2\hat{\boldsymbol{\Omega}}\,
\mathcal{P}_{x}^{*} \mathcal{P}_{y} F_a^{(i)*} F_b^{(i)} \right],
\label{eq:GammaXY_def}
\end{equation}
where $x\in\{0,a\}$ and $y\in\{0,b\}$ identify the Earth or pulsar contribution in the response of each member of the pair. This naturally extends the notation $\Gamma_{0,I}$ of Ref.~\cite{Liang:2021bct} for the Earth-Earth contribution, denoted here by $\Gamma_{ab}^{00,I}$.

\subsection{Finite-distance normalization}\label{subsec:finite_norm}

We first revisit the normalization convention for the tensor mode in GR. The usual normalization is one where $\widetilde{\Gamma}_{ab}^{\rm HD}$ is normalized to 1 in the case of the autocorrelation where pulsar $a = b$, and is normalized to 1/2 in the case of the cross correlation, where pulsar $a \neq b$ \cite{Cordes:2024oem, Agazie:2023}. The normalized HD correlation is thus written as 
\begin{equation}
    \begin{gathered}
        \widetilde{\Gamma}_{ab}^{\rm HD}(\xi_{a b})=\frac{3}{2} x \ln (x)-\frac{1}{4} x+\frac{1}{2}+\frac{1}{2} \delta_{a b}, \\
        x=\frac{1-\cos \xi_{a b}}{2} .
    \end{gathered}
\end{equation}
The $\delta_{ab}$ term introduces a discontinuity between the coaligned cross-correlation and the autocorrelation. This distinction arises from the pulsar terms, as we will see.

We consider the finite-distance factor in the GR limit, from Eq.~\eqref{eq:finite_distance_full}. For two distinct pulsars, the pulsar phases are rapidly oscillatory in the large-$fL$ limit. The pulsar terms, i.e.~those that contain $\mathcal{P}^*_{a}$ or $\mathcal{P}_{b}$, average away in the sky average integral, leaving only the Earth--Earth term, 1. This is the contribution that gives the usual HD curve, whose $\xi_{ab}\to0^+$ limit is 1/2.

For the autocorrelation, $a=b$, the pulsar--pulsar term becomes $ \mathcal{P}_{a}^{*}\mathcal{P}_{a}=1$, so the finite-distance factor becomes
\begin{equation}
\left(\mathcal{P}_{a}^{*}-1\right)
\left(\mathcal{P}_{a}-1\right)
=
2-\mathcal{P}_{a}^{*}-\mathcal{P}_{a}.
\end{equation}
The remaining two terms are still rapidly oscillatory and average away, but the non-oscillatory part of the sum is now 2 rather than 1. Thus the autocorrelation receives an Earth--Earth contribution and a pulsar--pulsar contribution that is equal in magnitude. This is where the $\delta_{ab}/2$ term comes from in the normalized HD curve, as emphasized in Ref.~\cite{Cordes:2024oem}.

However, this convention for normalizing the finite-distance GR tensor ORF, which we denote $\Gamma_{ab}^{\rm GR}$ to distinguish it from HD, is not able to properly account for pulsars that are very similar in distance. The Kronecker delta differentiates exactly identical pulsars from all of the distinct pulsar pairs, but the finite-distance response itself is not discontinuous in this way. The coherence of the pulsar--pulsar term is controlled by the relative phase between the two pulsar terms. Even for two distinct pulsars, this phase can be small if their distances are sufficiently similar. In that case, the pulsar--pulsar term does not necessarily behave like a rapidly oscillating factor, and the ORF may behave more like the autocorrelation.

\begin{figure}
    \centering
    \includegraphics[width=\linewidth]{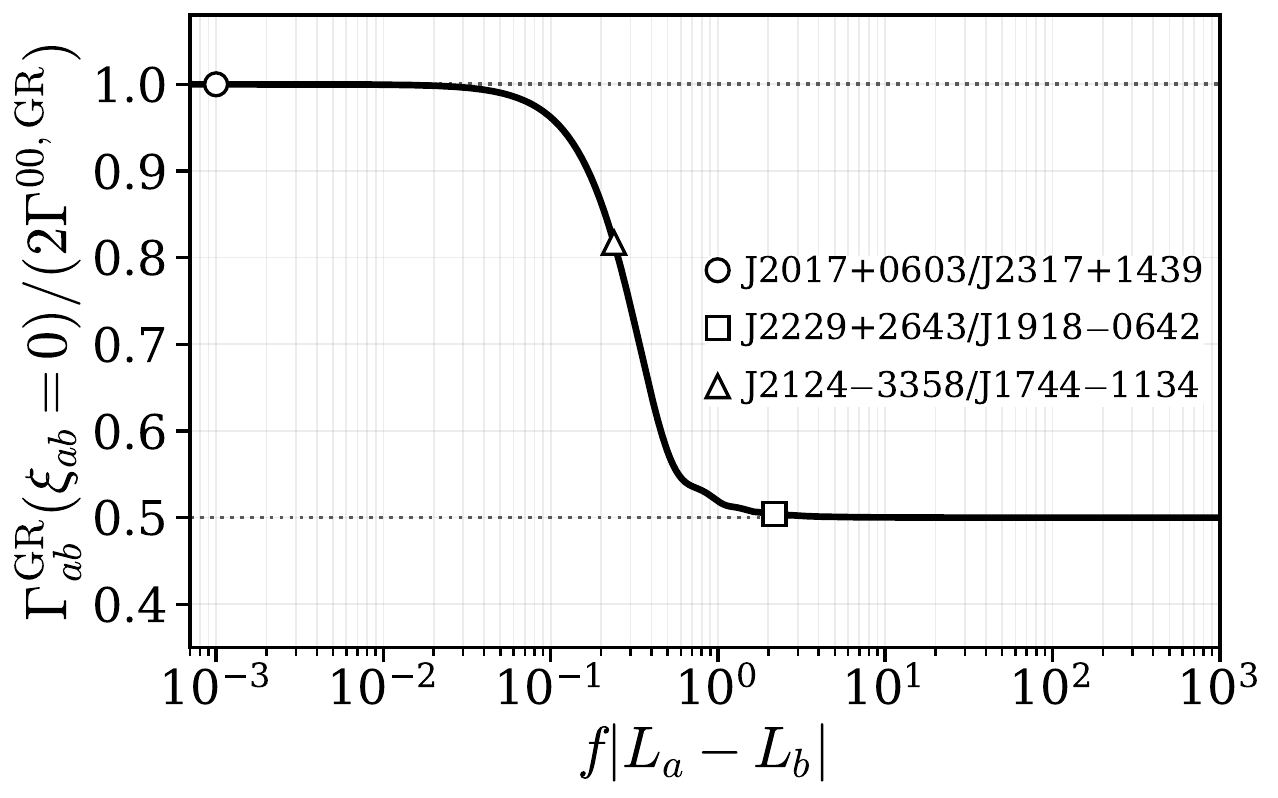}
    \caption{The tensor ORF in the massless GR limit at $\xi_{ab}=0$, normalized by the Earth--Earth contribution $2\Gamma^{00,T}$, in black. It is computed with $fL_a=500$ and $fL_b=fL_a+ f|L_a - L_b|$. The markers show selected pulsar pairs evaluated using their distances at a representative frequency $f=1/T_{\rm obs} = 1 / (16 \text{ yrs})$. The pair with equal distances is displayed at $f|L_a-L_b|=10^{-3}$ for visibility on the logarithmic scale.}
    \label{fig:tensor_distance_coherence}
\end{figure}
This can be seen in Fig.~\ref{fig:tensor_distance_coherence} for the tensor mode in the GR limit, where for small $f|L_a-L_b|$, the pulsar phases are coherent and the response approaches $1$, as in an autocorrelation. For large $f|L_a-L_b|$, the pulsar-dependent pieces decohere and the response approaches $1/2$. For distinct pulsars with nearly equal distances, partial coherence allows the coaligned response to take intermediate values between $1/2$ and $1$ instead of being fixed at $1/2$ whenever $a\neq b$. To retain this continuous dependence, we normalize $\Gamma^{\rm GR}_{ab}(\xi_{ab},f)$ by twice the tensor Earth--Earth response evaluated at $\xi_{ab}=0$, $2\Gamma^{00,\rm GR}$, which smoothly interpolates between the two limiting behaviors, as shown in  Fig.~\ref{fig:tensor_distance_coherence}. Here, we note that the Earth-Earth contribution of the ORF no longer has the subscript $ab$ because we have evaluated the ORF at $\xi_{ab} = 0$ and the pulsar distances no longer appear in the expression. 

With the tensor polarization convention adopted here, the Earth--Earth response in GR satisfies $\Gamma^{00, {\rm GR}} =2\pi/3$, and therefore
\begin{equation}\label{eq:beta_GR}
\beta_{\rm GR}
=
\frac{1}{2\Gamma^{00, {\rm GR}}}
=
\frac{3}{4\pi},
\end{equation}
which is also a standard normalization used in PTA ORF calculations
\cite{Anholm:2008wy,Chamberlin:2011ev,NANOGrav:2021ini}. Although the interchange between this normalization and the $\delta_{ab}$-based normalization is justified when pulsar terms are ignored, it can potentially yield ORFs that disagree in their overall amplitude when pulsar terms are kept. We find that keeping the constant $4\pi/3$ factor as the normalization explicitly preserves the continuous dependence on pulsar separation and distances when pulsar terms are kept.

This is relevant for real PTA observations. For example, pulsars J2017+0603 and J2317+1439 are both located at a distance of $1.57\ {\rm kpc}$, with uncertainties $\pm 0.31\ {\rm kpc}$ and $\pm 0.29\ {\rm kpc}$, respectively. Similarly, pulsars J2124$-$3358 and J1744$-$1134 are located $0.413\pm0.055\ {\rm kpc}$ and $0.4141\pm0.0093\ {\rm kpc}$ away. For this pair, the finite-distance normalization gives $\Gamma_{ab}^{\rm GR}(\xi_{ab}=0)/(2\Gamma^{00,{\rm GR}})\simeq0.82$ rather than $1/2$.

Each pulsar pair fixes the distances $L_a$ and $L_b$ and defines its own ORF as a function of angular separation. The correlation entering the analysis is evaluated at the actual observed separation $\xi_{ab}$, while the coaligned limit is obtained by varying the separation with both distances held fixed. This limit is relevant to the normalization of the entire ORF, regardless of the pair's observed separation. Thus, if we want to compare the ORF values for the different pulsar pairs, especially for a set that contains pairs with similar distances, a shared normalization that puts them on equal footing is necessary.

Our procedure for normalizing the ORF in GR elucidates an analogous normalization procedure for the effective ORF in massive gravity. As shown in Eq.~\eqref{eq:eff_orf_common_beta}, the effective ORF carries the normalization coefficient $\beta_T$. We therefore define $\beta_T$ by using the massive-gravity tensor Earth--Earth contribution,
\begin{equation}
\beta_T = \frac{1}{2\Gamma^{00,T}}.
\label{eq:beta_T_00}
\end{equation}
This quantity can be analytically computed:
\begin{equation}
\begin{aligned}
    \Gamma^{00,T} &= \frac{\pi}{4}\int_{0}^{\pi} d \theta \sin \theta \frac{(1-\cos^2 \theta)^2}{(1+v_g\cos\theta)^2} \\ &=
\frac{\pi}{3v_g^5}
\left[
6v_g - 4v_g^3 +3(v_g^2-1)\ln\left(\frac{1+v_g}{1-v_g}\right)
\right].
\end{aligned}
\label{eq:tensor_00_normalization}
\end{equation}
Its value in the massless limit is $2\pi/3$, recovering the standard HD normalization $\beta_T=3/(4\pi)$. In the opposite limit, as $m_g / \omega \to 1$ and $v_g\to0$, $\Gamma^{00,T}$ approaches the value $4\pi/15$, yielding $\beta_T = 15/(8\pi)$. Thus, the tensor Earth--Earth normalization remains finite across the full range of $v_g$.

We adopt the tensor Earth--Earth normalization to compare finite-distance responses across pulsar pairs and propagation regimes. The behavior of these responses depends on both $v_g$ and $\xi_{ab}$. In massive gravity, the pulsar phase contains the factor $1+v_g\hat{\boldsymbol{\Omega}}\cdot\hat{\boldsymbol{p}}_a$. As $v_g$ decreases, its angular oscillations become weaker, and the pulsar-dependent terms no longer necessarily average away over the sky. We examine the massive limit and, following the autocorrelation-first treatment of Ref.~\cite{Cordes:2024oem}, the massless autocorrelation limit obtained by first setting $b=a$, $\xi_{aa}=0$, and $L_b=L_a$ at finite $fL_a$, and then taking $v_g\to1$ at fixed $f$ and $L_a$.

\subsection{ORF expressions}\label{subsec:analytic_ORF}
To simplify our expressions, we define the angular projection factors for pulsar $a$ and $b$
\begin{equation}
\begin{aligned}
     \mu_a &\equiv \hat{\boldsymbol{\Omega}} \cdot \hat{\boldsymbol{p}}_a = \cos\theta , \\ 
    \mu_b &\equiv \hat{\boldsymbol{\Omega}} \cdot \hat{\boldsymbol{p}}_b = 
    \cos\theta\cos\xi_{ab}+ \sin\theta\cos\phi\sin\xi_{ab}.
\end{aligned}
\end{equation}
Substituting the polarization tensors into the general receiving functions in Eq.~\ref{eq:receiving} gives the following expressions
\begin{equation}\label{eq:helicity_receiving}
\begin{aligned}
\sum_{i=\pm2}F_a^{(i)*}F_b^{(i)} &= \frac{
(\cos\xi_{ab}-\mu_a\mu_b)^2
-\frac12(1-\mu_a^2)(1-\mu_b^2)}{4(1+v_g\mu_a)(1+v_g\mu_b)},\\[6pt]
\sum_{i=\pm1}F_a^{(i)*}F_b^{(i)} &= \frac{(1-v_g^2)\mu_a\mu_b(\cos\xi_{ab}-\mu_a\mu_b)}{2(1+v_g\mu_a)(1+v_g\mu_b)},\\[6pt]
F_a^{(0)*}F_b^{(0)} &= \frac{[(3-2v_g^2)\mu_a^2-1][(3-2v_g^2)\mu_b^2-1]}{24(1+v_g\mu_a)(1+v_g\mu_b)}.
\end{aligned}
\end{equation}
The intermediate contractions are provided in Appendix \ref{app:receiving_functions}. Factoring out the common denominators $(1 + v_g \mu_a)$ and $(1 + v_g \mu_b)$ and grouping them with the finite-distance factors of the ORF, since they naturally appear in the phases of the pulsar terms, gives us the angular kernels $\mathcal{N}_I$ for each polarization sector:
\begin{equation}
    \sum_{i\in I}F_a^{(i)*}F_b^{(i)} = \frac{\mathcal N_I}{(1+v_g\mu_a)(1+v_g\mu_b)}.
\end{equation}
The sector ORFs therefore take the common form
\begin{equation}\label{eq:GammaI_def}
\begin{aligned}
\Gamma_{ab}^{I}
&=
\operatorname{Re}\Bigg[
\int_0^\pi d\theta
\int_0^{2\pi} d\phi\,\sin\theta\,
\mathcal N_I(\theta,\phi,\xi_{ab};v_g)
\\
&\quad\times
\frac{e^{i2\pi fL_a(1+v_g\mu_a)}-1}
     {1+v_g\mu_a}
\frac{e^{-i2\pi fL_b(1+v_g\mu_b)}-1}
     {1+v_g\mu_b}
\Bigg],
\end{aligned}
\end{equation}
for $I\in\{T,V,S\}$. Note that because the receiving function sums are all real, the imaginary component of the integrand comes solely from the pulsar terms. We see that the tensor kernel is independent of $v_g$ because the helicity-$\pm2$ polarization tensors are purely spatial and transverse. Nevertheless, the tensor ORF has $v_g$ dependence through the denominators and the pulsar phases. The vector kernel contains a factor of $1 - v_g^2$, that converges to 0 when $v_g\to1$, causing the vector ORF to vanish in the GR limit. Like the tensor ORF, the scalar ORF remains finite in the GR limit. These ORFs are not analytically integrable in their full general forms, but in certain limits, they permit analytical forms.

\subsection{Massive limit} \label{sec:massive_limit}
We first analyze the massive limit, in which $m_g/\omega \to 1$, or equivalently $v_g \to 0$ when keeping $\omega$ fixed. In this limit, the angular dependence of the phases of the pulsar terms disappear because the oscillatory part of the pulsar term, $1+v_g \mu_a$ and $1+v_g\mu_b$, both approach 1. The pulsar terms thus approach the following forms, 
\begin{equation}
    \lim_{v_g\to0}\mathcal{P}_a^*=e^{i2\pi f L_a},
    \qquad
    \lim_{v_g\to0}\mathcal{P}_b=e^{-i2\pi f L_b}.
\end{equation}
Taking the real part of the finite-distance product in the ORF integrals, we get
\begin{equation}
\label{eq:pulsar_factor}
    \begin{aligned}
        \mathcal{R}_{ab}^{0} &\equiv \operatorname{Re}\!\left[ (e^{i2\pi fL_a}-1)(e^{-i2\pi fL_b}-1)\right] \\ &=\operatorname{Re}\!\left[ (2i e^{i\pi fL_a})(-2i e^{-i\pi fL_b}) \right]\sin(\pi fL_a)\sin(\pi fL_b) \\ &=4\cos[\pi f(L_a-L_b)] \sin(\pi fL_a)\sin(\pi fL_b).
    \end{aligned}
\end{equation}
Here we used the identities $e^{\pm ix}-1=\pm2i e^{\pm ix/2}\sin(x/2)$ and $\operatorname{Re}(e^{ix})=\cos x$. $\mathcal{R}_{ab}^0$ is periodic under $fL_a\to fL_a+n$ and $fL_b\to fL_b+m$, for integers $n$ and $m$. Because the pulsar-distance uncertainties relevant to PTAs generally span many cycles at nanohertz frequencies, the phases $2\pi fL_a$ and $2\pi fL_b$ are effectively unconstrained modulo $2\pi$. As a result, the factor can vary between pulsar pairs and may take either sign. Its maximum value of 4 occurs when both $fL_a$ and $fL_b$ are half-integers, and its minimum value of -1/2 occurs when the fractional parts of $fL_a$ and $fL_b$ are $1/6$ and $5/6$ in either order. Due to the lack of angular dependence of the function, it factors out of the angular integral and as a result, the angular dependence of the ORFs is determined entirely by $\mathcal{N}_T$, $\mathcal{N}_V$, and $\mathcal{N}_S$.

\begin{figure}
    \centering
    \includegraphics[width=\linewidth]{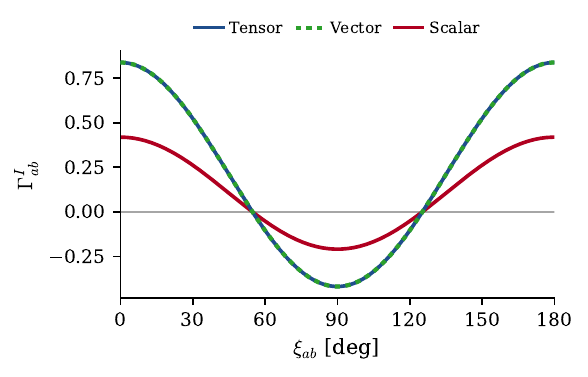}
    \caption{Angular correlations of the tensor, vector, and scalar sectors. The curves are not normalized and $\mathcal{R}_{ab}^0$ has been set to 1 to isolate their angular dependence. Note that the tensor and vector curves overlap, since they have identical angular correlations.}
    \label{fig:massive_limit}
\end{figure}

The angular integrals of these kernels give the sector ORFs in the massive limit:
\begin{equation}
\begin{aligned}
\Gamma^T_{ab}(v_g=0)
&=\Gamma^V_{ab}(v_g=0)
=\frac{4\pi}{15}R^0_{ab}P_2(\cos\xi_{ab}),
\\
\Gamma^S_{ab}(v_g=0)
&=\frac{2\pi}{15}R^0_{ab}P_2(\cos\xi_{ab}),
\end{aligned}
\end{equation}
where $P_2(\cos\xi_{ab})=(3\cos^2\xi_{ab}-1)/2$ is the Legendre polynomial of order 2. All three sectors share this quadrupolar dependence, with identical tensor and vector responses and a scalar response half their magnitude, as illustrated in Fig.~\ref{fig:massive_limit}.

We form the effective ORF by combining the three modes, setting the energy density ratios $\Omega_V/\Omega_T = 1$ and $\Omega_S/\Omega_T = 1/2$ and applying the normalization of Eq.~\eqref{eq:beta_T_00}. We find the massive limit to be
\begin{equation}
\widetilde{\Gamma}_{ab}^{\rm eff}(\xi_{ab},f) = \frac{5}{4}\mathcal{R}_{ab}^0 P_2(\cos\xi_{ab}) .
\label{eq:massive_limit_effective_orf}
\end{equation}
In this limit, the pulsar terms no longer average away, and $\mathcal{R}_{ab}^0$ modifies the normalized effective ORF, giving a possible enhancement, suppression, or a change in sign, depending on the pulsar phases. If the pulsar terms are instead neglected before computing the angular integral, this factor is replaced by 1 and the result becomes $5 P_2(\cos\xi_{ab})/4$.
\begin{figure}
    \centering
    \includegraphics[width=\linewidth]{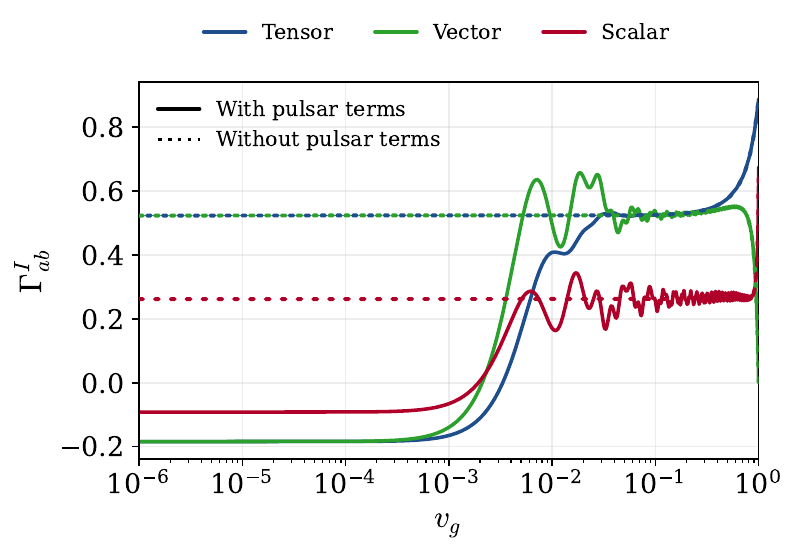}
    \caption{The ORFs for the tensor, vector, and scalar modes as a function of the group velocity $v_g$. We use representative distances $L_a = 500$ ly and $L_b = 2000$ ly, frequency $f = 1/(16\ \text{yrs})$, and an angular separation of $30^\circ$. }
    \label{fig:ORF_vg_massive}
\end{figure}

For the parameters in Fig.~\ref{fig:ORF_vg_massive}, $\mathcal R^0_{ab}=-0.3523$, so the pulsar terms reverse the signs of all three Earth–Earth responses and reduce their magnitudes to approximately $35\%$. We can see that even for large pulsar distances, increasing $m_g$ can prevent the pulsar-dependent terms from averaging away.

\subsection{Massless limit}
\label{sec:massless_limit}
We now examine the ORFs in the massless limit $m_g\to0$, or equivalently $v_g\to1$ keeping $\omega$ fixed. We first consider the ORFs obtained by dropping the pulsar terms to illustrate the necessity of the pulsar terms in this limit. In the massless limit, the tensor Earth-Earth contribution to the ORF remains finite and gives the usual
HD correlation
\begin{equation}
\begin{aligned}
    \lim_{v_g \to 1} \Gamma_{ab}^{00,T} &= 
\frac{\pi}{6}
\Bigg[
3+\cos\xi_{ab}\\ &
+6(1-\cos\xi_{ab})
\ln\left(\frac{1-\cos\xi_{ab}}{2}\right)
\Bigg].
\end{aligned}
\end{equation}
The autocorrelation of the massless tensor ORF with the pulsar terms is found to be
\begin{equation}
\Gamma^{\rm T}_{aa} = \frac{4\pi}{3}
\Bigg[ 1 
-\frac{3}{2(2\pi fL)^2}
+\frac{3\sin(4\pi fL)}{4(2\pi fL)^3}
\Bigg].
\label{eq:HD_auto_finite}
\end{equation}
We see that the autocorrelation value of $4\pi/3$ is only modified by small correction terms whose leading order scales as $(fL)^{-2}$ for large $fL$. If the tensor modes are the only modes considered in an analysis, then ignoring pulsar terms provides a very good approximation for cross-correlations between distinct, sufficiently separated pulsars. For autocorrelations, however, the pulsar--pulsar contribution must still be retained, producing the familiar factor of two in the large-$fL$ limit.

\begin{figure}
    \centering
    \includegraphics[width=\linewidth]{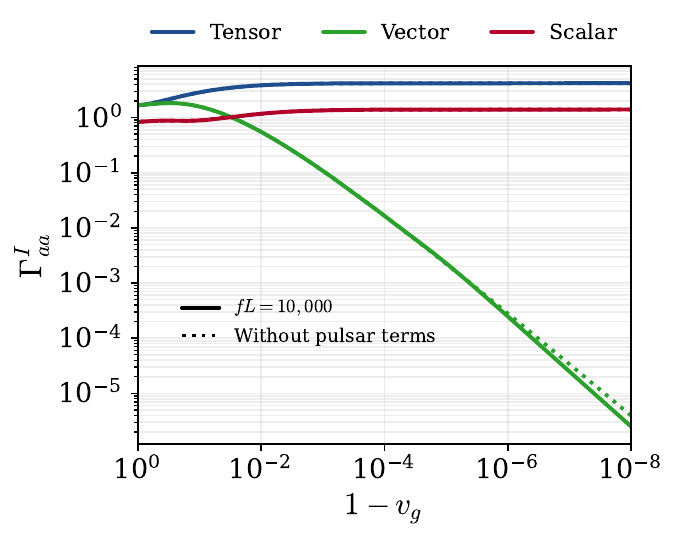}
    \caption{Autocorrelations of the tensor, vector, and scalar modes as functions of $1-v_g$. Solid curves show the full ORFs for $fL_a = fL_b =10^4$, while dotted curves show the ORF with pulsar terms removed and multiplied by two to match the autocorrelation normalization. For $v_g$ very close to 1, the vector ORF exhibits a noticeable difference in behavior depending on whether pulsar terms are retained.}
    \label{fig:auto_finite_distance_vg}
\end{figure}

To examine the Earth--Earth vector response near the massless limit, we factor out $1-v_g^2$, obtaining
\begin{equation}
\begin{aligned}
\lim_{v_g\to1}
\frac{\Gamma_{ab}^{00,V}}{1-v_g^2}
=
-\frac{2\pi}{3}\Bigg[
&3+4\cos\xi_{ab}\\
&+3\ln\!\left(
\frac{1-\cos\xi_{ab}}{2}
\right)
\Bigg].
\end{aligned}
\end{equation}
For fixed $0<\xi_{ab}\leq\pi$, this rescaled limit is finite, and
the Earth--Earth vector ORF therefore vanishes as $v_g\to1$.
The additional limit $\xi_{ab}\to0$ produces a logarithmic
divergence in the rescaled expression. To examine the coaligned
configuration, we first set $\xi_{ab}=0$ at finite $v_g$, obtaining
\begin{equation}
\begin{aligned}
\Gamma^{00,V}_{aa}
={}&\pi(1-v_g^2)
\int_0^\pi d\theta\,\sin\theta\,
\frac{\cos^2\theta\sin^2\theta}
     {(1+v_g\cos\theta)^2}
\\
={}&\!\frac{2\pi(1-v_g^2)}{3v_g^5}\!\!
\left[
2v_g^3-12v_g
+3(2-v_g^2)
\ln\!\left(\frac{1+v_g}{1-v_g}\right)\!
\right]\!.
\end{aligned}
\end{equation}
This expression also vanishes when $v_g \to 1$. The Cauchy--Schwarz bound $|\Gamma^{00,V}_{ab}|\leq\Gamma^{00,V}_{aa}$ extends this conclusion to all angular separations. Retaining the pulsar terms also makes the rescaled response regular at fixed finite pulsar distance, since
\begin{equation}\label{eq:remove_sing}
    \lim_{x\to0} \frac{e^{i2\pi fL x}-1}{x} = i 2\pi fL .
\end{equation}
The finite-distance pulsar phase therefore regulates the apparent singularity. For the vector ORF, the autocorrelation is
\begin{equation}
\begin{aligned}
\Gamma_{aa}^{V}
&=2\pi(1-v_g^2)
\int_{-1}^{1}d\mu_a\,
\frac{\mu_a^2(1-\mu_a^2)}{(1+v_g\mu_a)^2}
\\
&\quad\times
\left[
1-\cos\!\left(2\pi fL_a(1+v_g\mu_a)\right)
\right].
\end{aligned}
\end{equation}
Following Ref.~\cite{Cordes:2024oem}, we set $b=a$ in the full response before taking $v_g\to1$ at fixed $f$ and finite $L_a$, obtaining
\begin{equation}\label{eq:Gamma_V_limit}
\begin{aligned}
\lim_{v_g\to1}
\frac{\Gamma_{aa}^{V}}{1-v_g^2}
&=
4\pi\Bigg[
\gamma_E+\ln(4\pi fL)-\mathrm{Ci}(4\pi fL)
-\frac{7}{3}
\\
&+\frac{\sin(4\pi fL)}{2(2\pi fL)}
+\frac{2}{(2\pi fL)^2}
-\frac{\sin(4\pi fL)}{(2\pi fL)^3}
\Bigg],
\end{aligned}
\end{equation}
where $\gamma_E$ is the Euler--Mascheroni constant and $\mathrm{Ci}$ is the cosine integral function, defined by
\begin{equation}\label{eq:cosine_int_func}
    \mathrm{Ci}(x)
    \equiv
    -\int_x^\infty dt\,\frac{\cos t}{t}.
\end{equation}
$\mathrm{Ci}(x)$ tends to 0 as $x\to \infty$. The rescaled finite-distance autocorrelation therefore has a finite massless limit, and $\Gamma^V_{aa}=O(1-v_g^2)$. Its large-$fL_a$ behavior is obtained only after taking this limit, yielding the logarithmic dependence on $fL_a$. Although both the Earth-Earth term and the full expression with pulsar terms vanish in the massless limit due to the factor of $1 - v_g^2$, retaining the pulsar terms changes the rate at which the vector mode decouples. This can be observed in Fig.~\ref{fig:auto_finite_distance_vg}, where we see that the full vector autocorrelation with pulsar terms and the Earth-Earth term of the vector autocorrelation begin to noticeably differ when $1 - v_g < 10^{-5}$ for the chosen parameters. Nevertheless, the vector contribution vanishes in the massless limit because of the factor of $1-v_g^2$, even though the finite-distance effects can modify its behavior relative to the Earth--Earth approximation.

The same analysis can be done for the scalar sector. In the massless limit, the scalar kernel takes the form
\begin{equation}
\mathcal{N}_S(\theta,\phi,\xi_{ab};1) = \frac{(1-\mu_a^2)(1-\mu_b^2)}{24}.
\end{equation}
This numerator cancels the apparent singularities in the denominator 
\begin{equation}
\frac{\mathcal{N}_S(\theta,\phi,\xi_{ab};1)}{(1+\mu_a)(1+\mu_b)} = \frac{(1-\mu_a)(1-\mu_b)}{24}.
\end{equation}
When we compute the scalar ORF while ignoring pulsar terms, we obtain a rather simple expression,
\begin{equation}
\lim_{v_g\to1}\Gamma_{ab}^{00,S} = \frac{\pi}{18}(3+\cos\xi_{ab}).
\end{equation}
In general, the helicity-0 polarization contains both breathing and longitudinal contributions. Although individual longitudinal polarization components diverge as $m_g \to 0$, the inverse powers of $m_g$ cancel in the complete receiving function, leaving a finite breathing contribution and a longitudinal correction that is suppressed by $1 - v_g^2$, as shown in Appendix \ref{app:receiving_functions}. With pulsar terms retained, the latter vanishes uniformly over propagation directions as $v_g\to1$. The surviving response therefore yields the finite breathing-mode ORF \cite{Bernardo:2022rif, Chamberlin:2011ev, Hu:2022ujx}.

Although the scalar ORF can be analytically integrated when pulsar terms are included in the $v_g = 1$ limit (see Eq.~(25) of Ref.~\cite{Hu:2022ujx}), the resulting expression is rather difficult to parse, so we shall analyze the autocorrelation. Setting pulsar $b = a$, the scalar ORF autocorrelation is found to be
\begin{equation}\label{eq:scalar_auto_limit}
\Gamma^{\rm S}_{aa} = \frac{4\pi}{9}
\Bigg[1 -\frac{3}{2(2\pi fL)^2}
+\frac{3\sin(4\pi fL)}{4(2\pi fL)^3}
\Bigg],
\end{equation}
which is identical to the tensor autocorrelation in Eq.~\ref{eq:HD_auto_finite} divided by a factor of three. It is clear from this expression that, like for the tensor mode, the pulsar terms also do not appreciably affect the scalar ORF in the massless limit, with a leading correction proportional to $(fL)^{-2}$ for large $fL$.
\begin{figure}
    \centering
    \includegraphics[width=\linewidth]{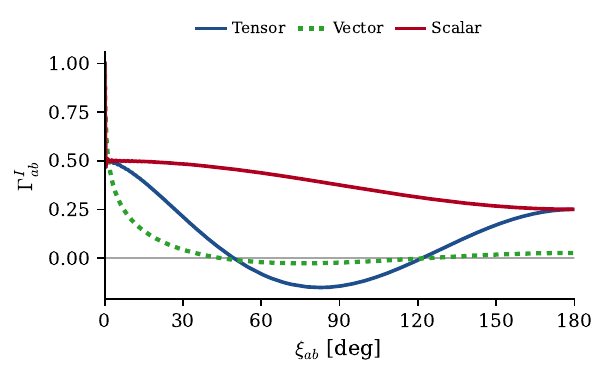}
    \caption{ORFs with pulsar terms in the massless limit, normalized to 1 at $\xi_{ab}=0$ to show their angular dependence. The $1 - v_g^2$ prefactor is removed from the vector ORF. We set $f=1/T_{\rm obs}$ with $T_{\rm obs}=16\,{\rm yr}$, and take $L_a=L_b=4500\,{\rm ly}$.}
    \label{fig:massless_limit_shapes}
\end{figure}

Fig.~\ref{fig:massless_limit_shapes} illustrates the massless limit ORF for each polarization sector. The tensor mode becomes the HD correlation, with small corrections due to the pulsar terms, the vector mode exhibits a pronounced peak near zero angular separation after the vanishing factor is removed, and the scalar mode approaches the breathing-mode correlation---a sum of the monopole and dipole correlations.

\begin{figure*}[t]
    \centering
    \includegraphics[width=\textwidth]{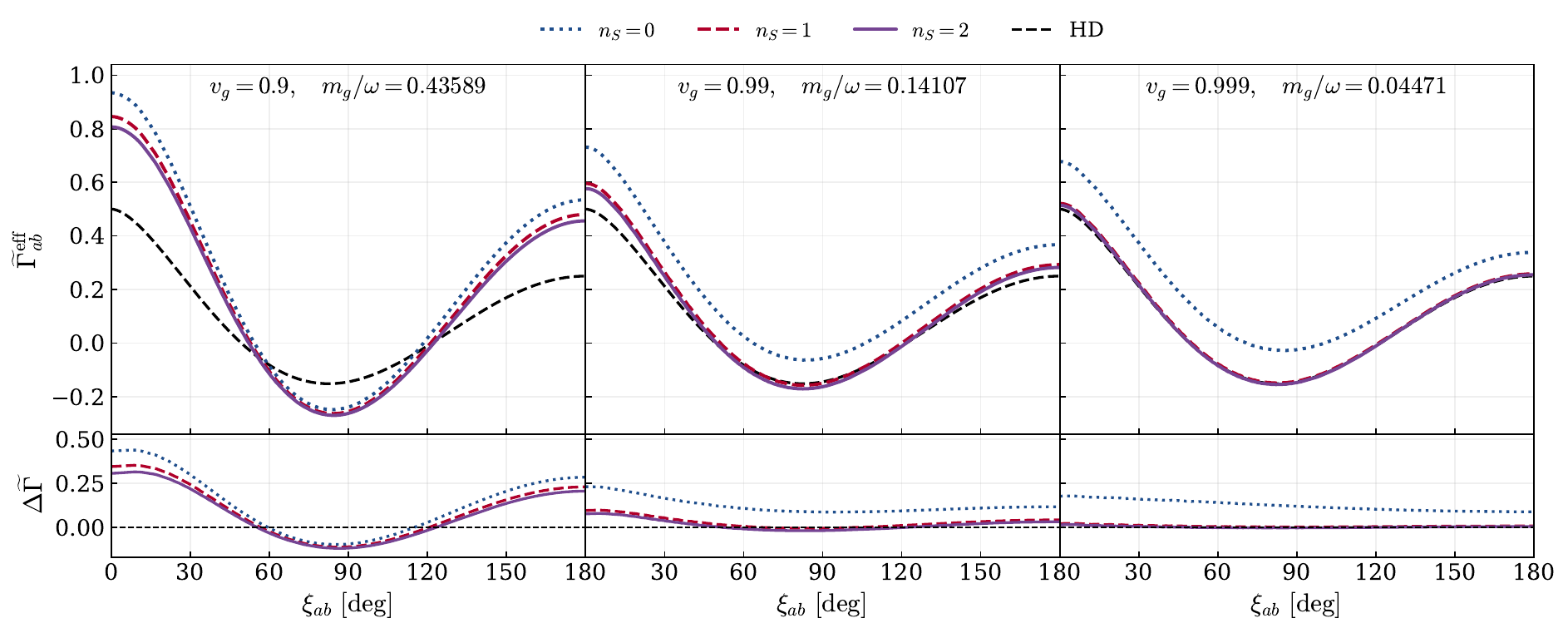}
    \caption{Effective ORFs for the scalar prescription $\Omega_S/\Omega_T=A_S(m_g/\omega)^{n_S}$, with $A_S=0.5$ and $n_S=0,1,2$. From left to right, the panels show $v_g=0.9,0.99,0.999$. The black dashed curve is the HD correlation. The lower panels show $\Delta\widetilde{\Gamma}=\widetilde{\Gamma}_{ab}^{\rm eff}-\Gamma_{\rm ab}^{\rm HD}$. We fix $\Omega_V/\Omega_T=1$, $f=1/(16.03\,{\rm yr})$, $L_a=500\,{\rm ly}$, and $L_b=2000\,{\rm ly}$.}
    \label{fig:scalar_suppression_orfs}
\end{figure*}

\section{Phenomenological scalar suppression}\label{sec:screen}
An important aspect that has not been adequately explored in earlier studies of ORFs in massive gravity is the treatment of the scalar mode as $m_g \to 0$. In this limit, the dispersion relation reduces to $\omega=|\boldsymbol{k}|$, and the effective ORF should recover the ORF in GR. However, in linearized massive gravity, the helicity-0 mode remains coupled to matter sources and can remain nonvanishing. In nonlinear completions of massive gravity, the helicity-0 contribution can be suppressed by the Vainshtein mechanism \cite{Vainshtein:1972sx, deRham:2010kj, deRham:2012fw, deRham:2014zqa}, for example. Taking Vainshtein screening as a qualitative motivation, we introduce a phenomenological prescription for suppressing the scalar contribution to the effective ORF.

We parametrize the scalar-to-tensor energy-density ratio as
\begin{equation}\label{eq:choice}
 \frac{\Omega_{S}}{\Omega_{T}} = A_S \left( \frac{m_g}{\omega} \right)^{n_S} ,
\end{equation}
where $n_S$ is the scalar suppression index that controls how quickly the scalar contribution decouples in the massless limit, and $A_S$ is the scalar amplitude parameter. We will show that $n_S>0$ is sufficient to make the scalar contribution vanish as $m_g/\omega\to0$.

\subsection{Convergence}\label{subsec:convergence}
We now prove that the effective ORF, with the scalar suppression, converges uniformly in $\xi_{ab}$ to the GR tensor ORF when $m_g \to 0$. We start with the scalar sector, whose contribution to the effective ORF is $2\beta_T A_S (m_g/\omega)^{n_S} \Gamma_{ab}^S$. The ORF is the real part of an inner product of the two pulsar response functions, as shown by Eq.~\ref{eq:orf_type}. The scalar ORF therefore satisfies the Cauchy-Bunyakovsky-Schwarz (CBS) inequality,
\begin{equation}
    \left|\Gamma_{ab}^{S}(\xi_{ab};f,L_a,L_b)\right|^2
    \leq
    \Gamma_{aa}^{S}(0;f,L_a,L_a)\Gamma_{bb}^{S}(0;f,L_b,L_b)
\end{equation}
for every $\xi_{ab}\in[0,\pi]$. The autocorrelations are nonnegative, since they are sky integrals of the squared moduli of the receiving function and finite-distance terms. At fixed $f$ and finite $L_r$, Eq.~\ref{eq:scalar_auto_limit}, together with the property $\sin x\leq x$ for $x>0$, gives
\begin{equation}
    0\leq\lim_{m_g\to0}\Gamma_{rr}^{S}
    \leq\frac{4\pi}{9},
    \qquad r\in\{a,b\}.
\end{equation}
The scalar autocorrelation is therefore bounded near the massless limit and is independent of $\xi_{ab}$. Since $\beta_T\to3/(4\pi)$, for fixed finite $A_S\geq0$ and $n_S>0$ we obtain
\begin{equation}
    \lim_{m_g\to0}
    \left|
        2\beta_T A_S
        \left(\frac{m_g}{\omega}\right)^{n_S}
        \Gamma_{ab}^{S}(\xi_{ab})
    \right|=0
\end{equation}
for every $\xi_{ab}\in[0,\pi]$. Since the CBS bound is independent of $\xi_{ab}$, and the scalar autocorrelations remain bounded as $m_g \to 0$, the scalar contribution converges uniformly to zero for all $\xi_{ab} \in [0,\pi]$.

More generally, the bound implies that the suppression does not need to be in the form of Eq.~\ref{eq:choice}. We see that any nonnegative scalar energy density ratio satisfying 
\begin{equation}\label{eq:general_constraint}
    \lim_{m_g\to0}\frac{\Omega_S}{\Omega_T}=0
\end{equation}
is able to suppress the scalar contribution uniformly in $\xi_{ab}$.

The same argument applies to the vector sector. At fixed $f$ and finite $L_a$ and $L_b$, the right-hand side of Eq.~\ref{eq:Gamma_V_limit} is finite, since it is not dependent on $v_g$. The factor $1-v_g^2$ therefore gives $\displaystyle\lim_{m_g\to0}\Gamma^V_{rr}=0$ for $r\in\{a,b\}$. The CBS inequality gives
\begin{equation}
    |\Gamma^V_{ab}(\xi_{ab})| \leq\sqrt{\Gamma^V_{aa}\Gamma^V_{bb}}
\end{equation}
for every $\xi_{ab}\in[0,\pi]$. Since the right-hand side of this inequality is independent of $\xi_{ab}$ and tends to zero when $m_g \to 0$, the vector ORF itself converges to zero in the massless limit. Additionally, since $\beta_T\to3/(4\pi)$, any bounded $\Omega_V /\Omega_T$ gives
\begin{equation}
    \lim_{m_g\to0} \left|
        \beta_T\frac{\Omega_V}{\Omega_T}
        \Gamma^V_{ab}
    \right|=0.
\end{equation}

For the tensor sector, we must show that the general tensor ORF uniformly converges to the GR tensor ORF as $m_g \to 0$, or equivalently, that the difference $\Gamma_{ab}^T - \Gamma_{ab}^{\rm GR}$ tends to zero for all $\xi_{ab}$. We see that Eq.~\ref{eq:GammaI_def} can be written using the pulsar factors defined in Eq.~\ref{eq:P_r}:
\begin{equation}
\begin{aligned}
    \Gamma_{ab}^{T}
    =\operatorname{Re}\Bigg[
    &\int_0^\pi d\theta\int_0^{2\pi}d\varphi\,
    \sin\theta\,\mathcal N_T
    \\
    &\times
    \frac{\mathcal P_a^*-1}{1+v_g\mu_a}
    \frac{\mathcal P_b-1}{1+v_g\mu_b}
    \Bigg].
\end{aligned}
\end{equation}
We recall that the tensor kernel $\mathcal N_T$ is independent of $v_g$. To examine the remaining dependence on $v_g$, we expand each exponential in its power series:
\begin{equation}
\begin{aligned}
    \frac{\mathcal P_r-1}{1+v_g\mu_r}
    &=
    \frac{
    \displaystyle\sum_{j=1}^{\infty}
    \frac{(-i2\pi fL_r)^j}{j!}
    (1+v_g\mu_r)^j
    }{1+v_g\mu_r}
    \\
    &=
    \sum_{j=1}^{\infty}
    \frac{(-i2\pi fL_r)^j}{j!}
    (1+v_g\mu_r)^{j-1}.
\end{aligned}
\end{equation}
This expression remains finite at $1+v_g\mu_r=0$. At fixed $f$ and finite $L_r$, the series and its first two derivatives with respect to $v_g$ converge uniformly, since $\mu_r$ is the dot product of two unit vectors, so $|\mu_r|\leq1$. We then Taylor expand the tensor ORF about $v_g=1$:
\begin{equation}
    \Gamma_{ab}^{T}(v_g)= 
    \Gamma_{ab}^{T}(1) +(v_g-1)
    \left.
    \frac{\partial\Gamma_{ab}^{T}}{\partial v_g}
    \right|_{v_g=1} + \!\mathcal O\bigl((v_g-1)^2\bigr).
\end{equation}
Since $\Gamma_{ab}^{T}(1)=\Gamma_{ab}^{\rm GR}$,
subtracting the GR ORF and using the Taylor expansion of $v_g$ with respect to $m_g / \omega$, 
\begin{equation}
    v_g-1
    =
    -\frac12\left(\frac{m_g}{\omega}\right)^2
    +\mathcal O\!\left(
    \left(\frac{m_g}{\omega}\right)^4
    \right),
\end{equation}
gives us
\begin{equation}
    \Gamma_{ab}^{T}(v_g)-\Gamma_{ab}^{\rm GR}
    =
    -\frac12\left(\frac{m_g}{\omega}\right)^2
    \left.
    \frac{\partial\Gamma_{ab}^{T}}{\partial v_g}
    \right|_{v_g=1}\!\!\!\!\! + \mathcal O\!\left(
    \left(\frac{m_g}{\omega}\right)^4
    \right).
\end{equation}
The derivative of $\Gamma_{ab}^T$ evaluated at $v_g=1$ is bounded independently of $\xi_{ab}$, and the same is true for the coefficient bounding the $\mathcal{O}((m_g/\omega)^4)$ terms.
We see that for sufficiently small $m_g/\omega$,
\begin{equation}
    \left|
    \Gamma_{ab}^{T}-\Gamma_{ab}^{\rm GR}
    \right|
    \leq C\left(\frac{m_g}{\omega}\right)^2,
\end{equation}
where $C$ is some finite constant. For the normalized tensor ORF, we write
\begin{equation}
\begin{aligned}
    \beta_T\Gamma_{ab}^{T}
    -\beta_{\rm GR}\Gamma_{ab}^{\rm GR}
    &=
    \beta_T
    \left(\Gamma_{ab}^{T}-\Gamma_{ab}^{\rm GR}\right)
    \\
    &\quad+
    (\beta_T-\beta_{\rm GR})\Gamma_{ab}^{\rm GR}.
\end{aligned}
\end{equation}
The CBS inequality also bounds the GR response by its finite autocorrelations from Eq.~\ref{eq:HD_auto_finite},
\begin{equation}
\begin{aligned}
    \left|
    \beta_T\Gamma_{ab}^{T}
    -\beta_{\rm GR}\Gamma_{ab}^{\rm GR}
    \right|
    &\leq
    |\beta_T|C\left(\frac{m_g}{\omega}\right)^2
    \\
    &\quad+
    |\beta_T-\beta_{\rm GR}|
    \sqrt{\Gamma_{aa}^{\rm GR}\Gamma_{bb}^{\rm GR}}.
\end{aligned}
\end{equation}
Since $\beta_T\to\beta_{\rm GR}=3/(4\pi)$, both
terms vanish as $m_g\to0$. The bound is independent
of $\xi_{ab}$, establishing uniform convergence of
the normalized tensor ORF to the GR ORF.

\subsection{Revisiting the massless limit}\label{subsec:revisit}
We now compare the rate at which the scalar contribution vanishes with the remaining tensor and vector departures from the GR tensor correlation. When the scalar mode is strictly unsuppressed, with the choice of $n_S = 0$ and nonzero $A_S$, the surviving scalar contribution adds a monopole-dipole angular dependence to the effective ORF. This can be seen in Fig.~\ref{fig:scalar_suppression_orfs}, where the unsuppressed effective ORF at $v_g= 0.999$, represented by the dotted blue curve in the rightmost panel, lies above HD. The corresponding residual in the lower panel approximately follows the scalar angular dependence $3+ \cos{\xi_{ab}}$. For $n_S > 0$, the scalar contribution vanishes as $m_g/\omega \to 0$. Already at $v_g = 0.999$, the choices $n_S = 1,2$ give effective ORFs close to HD, as shown by the dotted red and solid purple curves.
\begin{figure*}
    \centering
    \includegraphics[width=\textwidth]{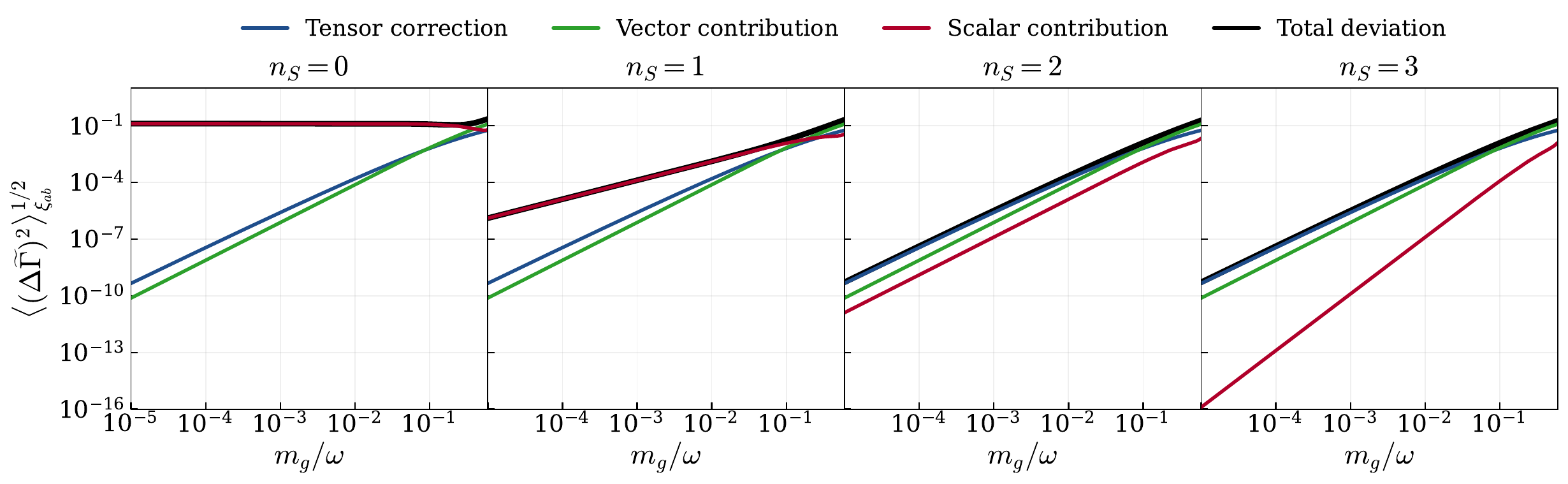}
    \caption{Convergence of the effective ORF toward its GR counterpart for $\Omega_S/\Omega_T=A_S(m_g/\omega)^{n_S}$, with $n_S=0,1,2,3$ from left to right. The blue curve shows the RMS of the difference between the massive and massless tensor ORFs; the green and red curves show the RMS of the vector and scalar contributions, respectively. The black curve is the RMS of the total deviation. We fix $A_S=0.5$, $\Omega_V/\Omega_T=1$, $f=1/(16.03\,{\rm yr})$, $L_a=500\,{\rm ly}$, and $L_b=2000\,{\rm ly}$.}
    \label{fig:scalar_deviation}
\end{figure*}

We wish to quantify to what extent the scalar contribution is responsible for the remaining deviation from the finite-distance GR tensor correlation. Although the tensor mode approaches $\Gamma_{ab}^{\rm GR}$ as $m_g \to 0$, at nonzero mass its response differs through the denominators $1 + v_g\mu_a$ and $1+ v_g\mu_b$ and the pulsar phases. Similarly, the vector mode can contribute at nonzero mass despite vanishing at the massless limit. We therefore quantify departures from the expected GR signal for the three sectors: 
\begin{equation}
    \begin{aligned}    
    \Delta\widetilde\Gamma^T_{ab}
    &=\beta_T\Gamma^T_{ab}
    -\widetilde\Gamma^{\rm GR}_{ab},
    \\
    \Delta\widetilde\Gamma^V_{ab}
    &=\beta_T\frac{\Omega_V}{\Omega_T}\Gamma^V_{ab},
    \\
    \Delta\widetilde\Gamma^S_{ab}
    &=2\beta_T A_S
    \left(\frac{m_g}{\omega}\right)^{n_S}
    \Gamma^S_{ab}.
    \end{aligned}
\end{equation}
The total deviation can be expressed as $\Delta\widetilde\Gamma^{\rm eff}_{ab}
=
\widetilde\Gamma^{\rm eff}_{ab}
-\widetilde\Gamma^{\rm GR}_{ab}
=
\Delta\widetilde\Gamma^T_{ab}
+\Delta\widetilde\Gamma^V_{ab}
+\Delta\widetilde\Gamma^S_{ab}.$
Then, integrating these quantities' squares over $\xi_{ab}$, we obtain the root mean squares (RMS)
\begin{equation}\label{eq:RMS}
    \langle (\Delta \widetilde{\Gamma}_{ab}^I)^2 \rangle^{1/2}_{\xi_{ab}}=\Bigg[ \frac{1}{2}\int_{0}^\pi d\xi_{ab} \sin \xi_{ab} \left| \Delta \widetilde{\Gamma}_{ab}^I \right|^2  \Bigg]^{1/2}.   
\end{equation}
This quantity is plotted in Fig.~\ref{fig:scalar_deviation}. For $n_S = 0,1$, the scalar contribution dominates remaining deviation from GR for sufficiently small $m_g / \omega$. The asymptotic slopes in Fig.~\ref{fig:scalar_deviation} follow from the convergence results established earlier.

For fixed $f$ and pulsar distances, the unnormalized tensor deviation scales as $\mathcal O((m_g/\omega)^2)$. The normalized tensor deviation, on the other hand, also contains the change in $\beta_T$. To find the further scaling of the  deviation, we expand $\beta_T=[2\Gamma^{00,T}]^{-1}$
\begin{equation}
\begin{aligned}
    \frac{\beta_T}{\beta_{\rm GR}}
    &=1+
    \left(\frac{m_g}{\omega}\right)^2
    \left[
    3\ln\!\left(\frac{2\omega}{m_g}\right)-4
    \right]
    \\
    &\quad+
    \mathcal O\!\left(
    \left(\frac{m_g}{\omega}\right)^4
    \ln^2\!\left(\frac{\omega}{m_g}\right)
    \right).
\end{aligned}
\end{equation}
Combining this with our results from Sec.~\ref{subsec:convergence}, the normalized tensor deviation scales in the massless regime as
\begin{equation}
\begin{aligned}
    \Delta\widetilde\Gamma_{ab}^{T}
    &=
    \beta_T
    \left(\Gamma_{ab}^{T}-\Gamma_{ab}^{\rm GR}\right)
    \\
    &\quad+
    \beta_{\rm GR}\left(\frac{\beta_T}{\beta_{\rm GR}}-1\right)
    \Gamma_{ab}^{\rm GR}
    \\
    &=3\left(\frac{m_g}{\omega}\right)^2
    \ln\left(\frac{\omega}{m_g}\right)
    \widetilde\Gamma_{ab}^{\rm GR}\\
    &\quad+\left(\frac{m_g}{\omega}\right)^2
    \Bigg[
    (3\ln 2-4)\widetilde\Gamma_{ab}^{\rm GR}\\
    &\qquad\qquad-
    \frac{\beta_{\rm GR}}{2}
    \left.\frac{\partial\Gamma_{ab}^{T}}{\partial v_g}\right|_{v_g=1}
    \Bigg]\\
    &\quad+\mathcal O\left[
    \left(\frac{m_g}{\omega}\right)^4
    \ln^2\left(\frac{\omega}{m_g}\right)
    \right].
\end{aligned}
\end{equation}
Since the logarithmic term, proportional to $(m_g/\omega)^2 \ln (\omega / m_g)$, converges more slowly than the term proportional to $(m_g/\omega)^2$ as $m_g/\omega\to 0$, it dominates the asymptotic behavior of the tensor deviation and is therefore the leading-order contribution. The vector response scales as $\mathcal{O}\left((m_g/\omega)^2\right)$ due to the factor $1 - v_g^2 = (m_g / \omega)^2$, while the remaining factor approaches a finite function of $\xi_{ab}$ in the massless limit. The scalar ORF approaches a finite limit, leaving its suppression factor to determine the leading scaling. Taking the angular RMS near $m_g \to 0$ therefore gives the following scalings:
\begin{equation}
    \left\langle
    |\Delta\widetilde\Gamma_{ab}^{I}|^2
    \right\rangle_{\xi_{ab}}^{1/2}
    \propto
    \begin{cases}
        \left(\displaystyle\frac{m_g}{\omega}\right)^2\ln\left(\displaystyle\frac{\omega}{m_g}\right), & I=T,\\
        \left(\displaystyle\frac{m_g}{\omega}\right)^2, & I=V,\\
        \left(\displaystyle\frac{m_g}{\omega}\right)^{n_S}, & I=S.
    \end{cases}
\end{equation}
 For $0 < n_S < 2$, the scalar contribution eventually dominates the remaining deviation. At exactly $n_S=2$, the scalar and vector contributions scale quadratically, while the tensor deviation contains an additional logarithmic factor. For $n_S>2$ the scalar contribution decays with an additional positive power of $m_g/\omega$, as seen in the rightmost panel of Fig.~\ref{fig:scalar_deviation}, where the scalar curve has a greater slope in $m_g/\omega$ and therefore is subdominant to the rest of the effective ORF's deviation from GR for low $m_g/\omega$.

\section{Conclusions and Discussion}\label{sec:conclusion}
In this paper, we analyzed the effective ORF for an SGWB in massive gravity, retaining finite-distance pulsar terms and suppressing the scalar sector. We showed that the pulsar terms remain important in the massive limit and produce pair-dependent amplitudes. We also introduced a normalization based on twice the tensor Earth--Earth response. In the massless limit, the vector ORF vanishes as $(m_g / \omega)^2$, whereas the scalar ORF approaches a finite, generally non-zero correlation, preventing recovery of the GR tensor correlation unless the scalar contribution is suppressed. We therefore introduced phenomenological suppression $\Omega_S/\Omega_T= A_S(m_g/\omega)^{n_S}$, with scalar amplitude $A_S \geq 0$ and suppression index $n_S>0$, and showed that it sufficiently decouples the scalar contribution to the effective ORF in order to recover the GR tensor ORF as $m_g\to0$. Together, these results provide a consistent treatment of the effective ORF for testing massive gravity with PTA observations near the massless limit.

This work may inspire several directions of research. For example, our phenomenological suppression formalism can be extended beyond PTAs to SGWB searches with ground- or space-based interferometers, where the long-wavelength regime dominates. ORFs for alternative polarizations have already been studied for ground-based detectors \cite{Nishizawa:2009bf, Callister:2017ocg}, and so a consistent treatment of the scalar mode in this context may be applied. Although ground-based searches have not yet detected an SGWB \cite{Virgo:2025aai}, their upper limits already constrain non-tensorial backgrounds \cite{LIGOScientific:2026mjf}. Extending our framework could allow us to find constraints using existing data and forecasts for future ground-based observations and for the Laser Interferometer Space Antenna \cite{LISACosmologyWorkingGroup:2022jok}.

A natural next step is to directly derive the suppression from the Vainshtein mechanism itself \cite{Vainshtein:1972sx}. In ghost-free massive gravity, the Vainshtein mechanism is associated with nonlinear interactions of the scalar mode \cite{deRham:2010kj, deRham:2012fw, deRham:2023ngf}. A theoretical prediction for the scalar energy density would require an explicit calculation that takes into account a specific model exhibiting the Vainshtein mechanism, and ties it to a specific source of GWs, such as a population of SMBHBs or a cosmological source \cite{Gumrukcuoglu:2012wt, Fujita:2018ehq, Choi:2023tun, Kenjale:2024rsc}.

An assumption we make is that $m_g$ used in Eq.~\ref{eq:dispersion_relation} is the mass scale that controls other manifestations of massive gravity, including static Yukawa forces and cosmological evolution. This is not necessarily the case in every nonlinear completion of massive gravity. For example, in extended quasidilaton massive gravity, the effective mass of GWs can differ from the graviton-mass parameter responsible for cosmic expansion \cite{Kahniashvili:2014wua, Gumrukcuoglu:2012wt}. In minimal quasidilaton massive gravity, massive tensor modes can instead have a high-momentum propagation speed equal to the speed of light \cite{DeFelice:2018vza}. A more complete treatment might include deriving the radiative dispersion relation within a nonlinear model.

We have also assumed the isotropy of the SGWB, which allows us to decompose Eq.~\ref{eq:two_point_z} as above. This assumption is not guaranteed: anisotropies may arise from discrete SMBHB sources, the large-scale distribution of matter, or features of a cosmological source \cite{Depta:2024ykq,Bravo:2025csu,Cusin:2025xle,Kuwahara:2024jiz,Li:2024lvt}. Anisotropic backgrounds are naturally described in harmonic space, where the multipole content of the SGWB sky distribution enters the timing-residual correlations through generalized, multipole-dependent ORFs that depend on the individual pulsar sky locations, not only on their angular separation \cite{Mingarelli:2013dsa,Gair:2014rwa,Allen:2024bnk,AnilKumar:2023yfw, Zhou:2026btr}. Fisher-matrix methods provide another way to work with real pulsar distributions \cite{Ali-Haimoud:2020ozu}. More recently, frequency-resolved optimal statistics have been developed to improve the characterization of SGWB anisotropies using PTA observations \cite{Gersbach:2025mhj}.

\begin{acknowledgments}
We thank Marcus Bosca, Neil Cornish, Murman Gurgenidze, Arthur Kosowsky, and Qiuyue Liang for useful discussions related to the paper. C.C.~and T.K.~acknowledge support from the NASA Astrophysics Theory Program (ATP) Award 80NSSC22K0825 and the National Science Foundation (NSF) Astronomy and Astrophysics Research Grants (AAG) Award AST2408411. T.K.~and B.P.~acknowledge partial support from the NSF AAG Award AST2307698. T.K.~acknowledges partial support from the Shota Rustaveli National Science Foundation of Georgia, FR24-901.
\end{acknowledgments}

\section*{Data Availability}
The source code used to reproduce the figures and numerical results in this work is publicly available on Zenodo \cite{Choi:2026git}.

\appendix 

\section{Receiving functions}
\label{app:receiving_functions}

We derive the receiving functions entering the
ORFs. We use $\mu_r=\hat{\boldsymbol\Omega}\cdot\hat{\boldsymbol p}_r$
throughout. We remind ourselves that the receiving function is
\begin{equation}
 F_r^{(i)}
 =-\frac{\hat p_r^\mu\hat p_r^\nu\epsilon_{\mu\nu}^{(i)}}
 {2[1+v_g\mu_r]}
 +\hat p_r^\mu\epsilon_{0\mu}^{(i)}-\frac12\epsilon_{00}^{(i)}
 [1+v_g\mu_r].
 \label{eq:app_receiving_general}
\end{equation}
For the polarization vectors given in the main text, we express them as
\begin{equation}
\begin{aligned}   
 \epsilon_0^\pm&=0,
 \\ \epsilon_0^0&=\frac{v_g}{\sqrt{1-v_g^2}}, \\ \hat p_r^\mu\epsilon_\mu^0
 &=\frac{v_g+\mu_r}
 {\sqrt{1-v_g^2}}.
 \label{eq:app_vector_contractions}
\end{aligned}
\end{equation}
The transverse vectors satisfy following identities:
\begin{equation}
\begin{aligned}\label{eq:identities}
 2\left|\hat p_r^\mu\epsilon_\mu^+\right|^2
 &=1-(\mu_r)^2,\\
 \sum_{\pm}
 (\hat p_a^\mu\epsilon_\mu^\pm)^*
 (\hat p_b^\nu\epsilon_\nu^\pm)
 &=\hat{\boldsymbol p}_a\cdot\hat{\boldsymbol p}_b
 -(\mu_a)
  (\mu_b).
\end{aligned}
\end{equation}
Since $\epsilon_\mu^-=(\epsilon_\mu^+)^*$, the sum in Eq.~\eqref{eq:identities} is twice the real part of its $+$ term.

\subsection{Tensor sector}
For $\epsilon_{\mu\nu}^{(\pm2)}=\epsilon_\mu^\pm\epsilon_\nu^\pm$, the time components vanish. Eq.~\eqref{eq:app_receiving_general} therefore reduces to
\begin{equation}
 F_r^{(\pm2)}
 =-\frac{(\hat p_r^\mu\epsilon_\mu^\pm)^2}
 {2[1+v_g\mu_r]}.
 \label{eq:app_tensor_response}
\end{equation}
The numerator of the helicity sum is
\begin{equation}
\begin{aligned}    
 &\sum_{\pm}
 \bigl[(\hat p_a^\mu\epsilon_\mu^\pm)^*\bigr]^2
 (\hat p_b^\nu\epsilon_\nu^\pm)^2 = 2\operatorname{Re}\!\left\{
 \left[(\hat p_a^\mu\epsilon_\mu^+)^*
       (\hat p_b^\nu\epsilon_\nu^+)\right]^2\right\}\\
 &\quad=4\left[\operatorname{Re}\!
 \left\{(\hat p_a^\mu\epsilon_\mu^+)^*
       (\hat p_b^\nu\epsilon_\nu^+)\right\}\right]^2
 -2|\hat p_a^\mu\epsilon_\mu^+|^2
    |\hat p_b^\nu\epsilon_\nu^+|^2,
 \label{eq:app_tensor_numerator}
\end{aligned}
\end{equation}
where we have used the identity $2 \operatorname{Re}(z^2) = 4[\operatorname{Re}(z)]^2 - 2 |z|^2 $. Using Eq.~\eqref{eq:identities}, together with $\hat{\boldsymbol p}_a\cdot\hat{\boldsymbol p}_b=\cos\xi_{ab}$, gives 
\begin{equation}
\begin{aligned}
\operatorname{Re}\!\left[
(\hat p_a^\mu\epsilon_\mu^+)^*
(\hat p_b^\nu\epsilon_\nu^+)
\right]
&=\frac{\cos\xi_{ab}-\mu_a\mu_b}{2},
\\
|\hat p_r^\mu\epsilon_\mu^+|^2
&=\frac{1-\mu_r^2}{2}.
\end{aligned}
\end{equation}
Substituting these into the tensor helicity sum yields
\begin{equation}
    \begin{aligned}
        \sum_{\pm}F_a^{(\pm2)*}F_b^{(\pm2)}\!&=\frac{4\left[\frac{\cos\xi_{ab}-\mu_a\mu_b}{2}\right]^2 -2\left[\frac{1-\mu_a^2}{2}\right] \left[\frac{1-\mu_b^2}{2}\right]}{4(1+v_g\mu_a)(1+v_g\mu_b)} \\ &=\frac{(\cos\xi_{ab}-\mu_a\mu_b)^2\! -\!\frac12(1-\mu_a^2)(1-\mu_b^2)}{4(1+v_g\mu_a)(1+v_g\mu_b)}.
    \end{aligned}
\label{eq:app_tensor_sum}
\end{equation}

\subsection{Vector sector}
For vector polarization tensors, $\epsilon_{\mu\nu}^{(\pm1)} =(\epsilon_\mu^\pm\epsilon_\nu^0 +\epsilon_\mu^0\epsilon_\nu^\pm)/\sqrt2$, the contractions are
\begin{equation}
\begin{aligned}   
 \hat p_r^\mu\hat p_r^\nu\epsilon_{\mu\nu}^{(\pm1)}
 &=\frac{\sqrt2[v_g+
 \mu_r]}
 {\sqrt{1-v_g^2}}\hat p_r^\mu\epsilon_\mu^\pm,
 \nonumber\\
 \hat p_r^\mu\epsilon_{0\mu}^{(\pm1)}
 &=\frac{v_g}{\sqrt{2(1-v_g^2)}}
 \hat p_r^\mu\epsilon_\mu^\pm, \\ \epsilon_{00}^{(\pm1)}&=0.
 \label{eq:app_vector_tensor_contractions}
\end{aligned}
\end{equation}
Substituting into Eq.~\eqref{eq:app_receiving_general}, we obtain
\begin{equation}
\begin{aligned}
F_r^{(\pm1)}
={}&
-\frac{v_g+\mu_r}
{\sqrt{2(1-v_g^2)}
 [1+v_g\mu_r]}
\hat p_r^\mu\epsilon_\mu^\pm
\\
&+\frac{v_g}{\sqrt{2(1-v_g^2)}}
\hat p_r^\mu\epsilon_\mu^\pm
-\frac12(0)
[1+v_g\mu_r]
\\
={}&
\frac{-(v_g+\mu_r) + v_g[1+v_g\mu_r]}
{\sqrt{2(1-v_g^2)}
 [1+v_g\mu_r]}
\hat p_r^\mu\epsilon_\mu^\pm
\\
={}&
-\frac{\sqrt{1-v_g^2}
(\mu_r)}
{\sqrt2[1+v_g\mu_r]}
\hat p_r^\mu\epsilon_\mu^\pm.
\end{aligned}
\label{eq:app_vector_response}
\end{equation}
We see that the terms in the numerator combine to form $-(1-v_g^2)\mu_r$, cancelling the inverse power of $\sqrt{1-v_g^2}$. Multiplying the vector receiving functions and summing over the
two helicities gives
\begin{equation}
\begin{aligned}
\sum_{\pm}(F_a^{(\pm1)})^*F_b^{(\pm1)}
={}&
\frac{
(1-v_g^2)
(\mu_a)
(\mu_b)}
{2[1+v_g\mu_a]
  [1+v_g\mu_b]}
\\
&\times
\sum_{\pm}
(\hat p_a^\mu\epsilon_\mu^\pm)^*
(\hat p_b^\nu\epsilon_\nu^\pm)
\\
={}&
\frac{
(1-v_g^2)
(\mu_a)
(\mu_b)}
{2[1+v_g\mu_a]
  [1+v_g\mu_b]}
\\
&\times
\left[
\cos\xi_{ab}
-(\mu_a)
 (\mu_b)
\right],
\end{aligned}
\label{eq:app_vector_sum}
\end{equation}
where the second equality follows from Eq.~\eqref{eq:identities}.

\subsection{Scalar sector}
For $\epsilon_{\mu\nu}^{(0)} =(\epsilon_\mu^+\epsilon_\nu^- +\epsilon_\mu^-\epsilon_\nu^+ -2\epsilon_\mu^0\epsilon_\nu^0)/\sqrt6$, all three terms of Eq.~\eqref{eq:app_receiving_general} contribute. The contractions are
\begin{align}
 \hat p_r^\mu\hat p_r^\nu\epsilon_{\mu\nu}^{(0)}
 &=\frac1{\sqrt6}\left[
 1-(\hat{\boldsymbol\Omega}\cdot\hat{\boldsymbol p}_r)^2 -\frac{2[v_g+\hat{\boldsymbol\Omega}\cdot\hat{\boldsymbol p}_r]^2}{1-v_g^2}\right],
 \nonumber\\
 \hat p_r^\mu\epsilon_{0\mu}^{(0)}
 &=-\frac{2v_g[v_g+
 \hat{\boldsymbol\Omega}\cdot\hat{\boldsymbol p}_r]}
 {\sqrt6(1-v_g^2)},
 \nonumber\\
 \epsilon_{00}^{(0)}
 &=-\frac{2v_g^2}{\sqrt6(1-v_g^2)}.
 \label{eq:app_scalar_contractions}
\end{align}
After substituting into Eq.~\ref{eq:app_receiving_general}, we get
\begin{equation}
\begin{aligned}
F_r^{(0)}
={}&
-\frac{1}{2\sqrt6(1+v_g\mu_r)}
\left[
1-\mu_r^2-\frac{2(v_g+\mu_r)^2}{1-v_g^2}
\right]
\\
&-\frac{2v_g(v_g+\mu_r)}{\sqrt6(1-v_g^2)}
+\frac{v_g^2(1+v_g\mu_r)}{\sqrt6(1-v_g^2)}
\\
={}&
-\frac{1-\mu_r^2}{2\sqrt6(1+v_g\mu_r)}
+\frac{(v_g+\mu_r)^2}
{\sqrt6(1-v_g^2)(1+v_g\mu_r)}
\\
&-\frac{2v_g(v_g+\mu_r)}{\sqrt6(1-v_g^2)}
+\frac{v_g^2(1+v_g\mu_r)}{\sqrt6(1-v_g^2)}.
\end{aligned}
\end{equation}
Combining the last three terms with a common denominator of $\sqrt6(1-v_g^2)(1+v_g\mu_r)$ gives us
\begin{equation}
\begin{aligned}
& F_r^{(0)}
=-\frac{1-\mu_r^2}{2\sqrt6(1+v_g\mu_r)} \\ &+\frac{
(v_g+\mu_r)^2 -2v_g(v_g+\mu_r)(1+v_g\mu_r)
+v_g^2(1+v_g\mu_r)^2}{\sqrt6(1-v_g^2)(1+v_g\mu_r)}.
\end{aligned}
\end{equation}
The numerator of the second term turns out to be a perfect square:
\begin{equation}
\begin{aligned}
&(v_g+\mu_r)^2
-2v_g(v_g+\mu_r)(1+v_g\mu_r)
+v_g^2(1+v_g\mu_r)^2
\\
&\qquad=
\left[(v_g+\mu_r)-v_g(1+v_g\mu_r)\right]^2
\\
&\qquad=
\left[(1-v_g^2)\mu_r\right]^2.
\end{aligned}
\end{equation}
One factor of $1-v_g^2$ therefore cancels, yielding
\begin{equation}
\begin{aligned}
F_r^{(0)}
&=-\frac{1-\mu_r^2}{2\sqrt6(1+v_g\mu_r)}
+\frac{(1-v_g^2)\mu_r^2}{\sqrt6(1+v_g\mu_r)}
\\
&=\frac{-(1-\mu_r^2)+2(1-v_g^2)\mu_r^2}
{2\sqrt6(1+v_g\mu_r)}
\\
&=\frac{(3-2v_g^2)\mu_r^2-1}
{2\sqrt6(1+v_g\mu_r)}.
\end{aligned}
\label{eq:app_scalar_response}
\end{equation}
The scalar receiving function is real, so the receiving function product is simply a square
\begin{equation}
\begin{aligned}
F_a^{(0)*}F_b^{(0)}
&=\frac{1}{24[1+v_g\mu_a][1+v_g\mu_b]}
\\
&\quad\times[(3-2v_g^2)\mu_a^2-1]
\\
&\quad\times[(3-2v_g^2)\mu_b^2-1].
\end{aligned}
\label{eq:app_scalar_product}
\end{equation}

\bibliographystyle{apsrev4-2}
\bibliography{suppression_theory_ref}

\clearpage
\end{document}